\documentclass[%
 reprint,
 amsmath,amssymb,
 aps,
]{revtex4-2}

\usepackage{float,subfigure}
\usepackage{cancel}
\usepackage{tikz}
\usepackage[export]{adjustbox}

\begin{document}

\title{Jumpstarting (elliptic) symbol integrations for loop integrals}
\author{Song He$^{1,2,3}$\footnote{songhe@itp.ac.cn}, Yichao Tang$^{1,4}$\footnote{tangyichao@itp.ac.cn}}
\affiliation{
$^{1}$CAS Key Laboratory of Theoretical Physics, Institute of Theoretical Physics, Chinese Academy of Sciences, Beijing 100190, China \\
$^{2}$School of Fundamental Physics and Mathematical Sciences, Hangzhou Institute for Advanced Study;\\
International Centre for Theoretical Physics Asia-Pacific, Beijing/Hangzhou, China\\
$^{3}$Peng Huanwu Center for Fundamental Theory, Hefei, Anhui 230026, P. R. China\\
$^{4}$School of Physical Sciences, University of Chinese Academy of Sciences, No.19A Yuquan Road, Beijing 100049, China
}\date{\today}

\begin{abstract}
We derive an algorithm for computing the total differentials of multi-loop integrals expressed as one-fold integrals of multiple polylogarithms, which can involve square roots of polynomials up to degree four and may evaluate to (elliptic) multiple polylogarithms ((e)MPL). This gives simple algebraic rules for computing the $(W{-}1, 1)$-coproduct of the resulting weight-$W$ functions up to period terms, and iterating it gives the symbol without actually performing any integration. In particular, our algorithm generalizes existing MPL integration rules and sidesteps the complicated rationalization procedure in the presence of square roots.
We apply our algorithm to 
conformal double-$D$-gon integrals in $D$ dimensions with generic kinematics and possibly massive circumferential propagators.
We directly compute, for the first time, the total differential and symbol (up to period terms) of the $D{=}3$ double-triangle and the $D{=}4$ double-box, which in the special case with massless propagators represent the first appearance of eMPL functions in (two-loop) scattering amplitudes of ${\cal N}{=}6$ Chern-Simons-matter theory and ${\cal N}{=}4$ super-Yang-Mills, respectively.
\end{abstract}
\maketitle
\section{Introduction}
The key for precise predictions in perturbative Quantum Field Theory (QFT) lies in the analytic computation of Feynman integrals, which often reveals rich and unexpected structures of QFT itself. Recent years have witnessed enormous progress in computing Feynman integrals, scattering amplitudes, {\it etc.}, which evaluate to the simplest class of functions, multiple polylogarithms (MPL)~\cite{Chen:1977oja,Goncharov1995GeometryOC,Goncharov:1998kja,Remiddi:1999ew,Borwein:1999js,Moch:2001zr}. At least for simple kinematics, a systematic method to compute (dimensionally regularized) Feynman integrals is via differential equations~\cite{Kotikov:1990kg,Kotikov:1991pm,Remiddi:1997ny,Gehrmann:1999as,Henn:2013pwa}. For complicated kinematics, other than direct integration~\cite{Bourjaily:2018aeq,Bourjaily:2019jrk,Bourjaily:2019vby,Bourjaily:2021lnz,Panzer:2014caa,Duhr:2019tlz,Gint,Caron-Huot:2011zgw,He:2020uxy,He:2020lcu}, it is often possible to bootstrap a Feynman integral~\cite{Chicherin:2017dob,Henn:2018cdp,He:2021esx,He:2021eec,He:2022ctv} once we have control over its analytic structure.

The analytic structure of MPLs is well understood due to powerful mathematical tools such as the {\it symbol} and the more general {\it coproduct}~\cite{Goncharov:2005sla,Goncharov:2010jf,Spradlin:2011wp,Duhr:2011zq,Duhr:2012fh} which manifests singularity structures and trivializes function identities. In essence, the symbol maps a complicated MPL Feynman integral to a tensor of simple {\it symbol letters} $\log f_i$, with $f_i$ algebraic functions of kinematics. 
More generally, Feynman integrals evaluate to more complicated functions (see~\cite{Bourjaily:2022bwx} and references therein), the simplest case involving elliptic multiple polylogarithms (eMPL)~\cite{Laporta:2004rb,brown2011multiple,Muller-Stach:2012tgz,Adams:2013nia,Bloch:2013tra,Adams:2014vja,Adams:2015gva,Adams:2015ydq,Adams:2016xah,Adams:2017tga,Adams:2017ejb,Bogner:2017vim,Broedel:2017kkb,Broedel:2017siw,Adams:2018yfj,Broedel:2018iwv,Broedel:2018qkq,Honemann:2018mrb,Broedel:2019hyg,Bogner:2019lfa,Duhr:2019rrs,Walden:2020odh,Weinzierl:2020fyx,Kristensson:2021ani,Wilhelm:2022wow,Giroux:2022wav,Morales:2022csr}
, for which one can define the symbol as well~\cite{Broedel:2018iwv}. For example, the symbol of eMPL double-box integrals contributing to two-loop ten-point amplitudes in ${\cal N}=4$ super-Yang-Mills theory (SYM) has been computed~\cite{Kristensson:2021ani,Wilhelm:2022wow}, which exhibits remarkably simple structures. For MPLs and eMPLs alike, the symbol is defined recursively by total differentials (or, equivalently, $(W{-}1,1)$-coproducts):
\begin{equation}\label{eq:symboldef}
    {\rm d}{\cal I}=\sum_\alpha {\cal I}_\alpha\,{\rm d}w_\alpha\implies\mathcal S({\cal I})=\sum_\alpha\mathcal S({\cal I}_\alpha)\otimes w_\alpha,
\end{equation}
where ${\cal I}$ and ${\cal I}_\alpha$ have \emph{transcendental weight} $W$ and $(W{-}1)$, and
the symbol letters $w_\alpha$ are one-fold integrals of rational functions over genus-one (elliptic) curves and genus-zero degenerations (\emph{i.e.}, MPL letters $\log f_i$).

In this Letter, we propose an algorithm for the direct computation of the symbol of MPLs and eMPLs expressed as one-fold integrals ${\cal I}=\int F(t)\,{\rm d}t$ of MPLs $F(t)$,
which applies to a large class of
Feynman integrals~\footnote{We mainly consider finite integrals in integer dimensions, but the method applies to each order in $\epsilon$ to dimensionally regularized integrals, and to integrals with mass regulators. It can also be used for the direct integration of amplitudes, Wilson loops~\cite{Caron-Huot:2011dec}, \emph{etc.}}. Algorithms exist~\cite{Caron-Huot:2011dec} that compute ${\rm d}\mathcal I$ and iteratively $\mathcal S(\mathcal I)$ in terms of $\mathcal S(F(t))$, as long as singularities of $F(t)$ involve linear factors of $t$ only.
However, it was previously unknown how to perform such {\it symbol integrations} when singularities of $F(t)$ involve square roots of polynomials of $t$.
We take an important step in solving this long-standing problem by deriving algebraic rules for ${\rm d}{\cal I}$ and iteratively ${\cal S}({\cal I})$, given ${\rm d}F(t)$. Our method sidesteps rationalization and gives the MPL symbol in the presence of square roots of quadratic polynomials. In the presence of square roots of cubics/quartics, it computes the eMPL symbol up to \emph{period terms} where $w_\alpha{=}\tau$, the modular parameter of the elliptic curve. The restriction to non-period terms, which is also the goal of elliptic symbol bootstrap~\cite{Morales:2022csr}, is often convenient in the study of eMPL symbols, since the period terms can be reconstructed via the \emph{symbol prime}~\cite{Wilhelm:2022wow}.

We apply our new method to the symbol integration of an important class of conformal integrals, double-$D$-gons in $D$ dimensions~\cite{Paulos:2012nu,Nandan:2013ip}, which are weight-$D$ and can be expressed as one-fold integrals of deformed $2(D{-}1)$-gons~\cite{Spradlin:2011wp,Abreu:2017ptx, Arkani-Hamed:2017ahv,Herrmann:2019upk,Bourjaily:2019exo}. We compute their total differentials, or $(D{-}1, 1)$-coproducts, even in the presence of massive circumferential propagators. The most general double-triangle in $D{=}3$ and double-box in $D{=}4$ depend on $6$ and $15$ conformal cross-ratios respectively, and up to period terms we obtain all last entries $w_\alpha$ as well as the (symbol of) accompanying weight-$(D{-}1)$ integrals which evaluate to MPL. In the special case with massless propagators, they reduce to the first eMPL contributions to scattering amplitudes in ABJM and SYM theory, respectively. For higher $D$, these weight-$(D{-}1)$ integrals involve elliptic or even higher-genus curves, and we leave their explicit computation to future work.  

\section{Deriving rules for (elliptic) symbol integrations}
\subsection{$2$-forms and MPL symbol integrations without rationalization}
Before studying elliptic integrals, let us first derive the symbol integration for MPL functions. We use $\delta:=\delta u\,\partial_u$ for the differential with respect to variables $\{u\}$ parametrizing the kinematic space $\mathcal K$, to distinguish it from the differential ${\rm d}:={\rm d}t\,\partial_t$ with respect to the integration variable $t$. It is helpful to consider the big space $\mathcal M$ parametrized by $\{u,t\}$ with total differential operator $D=\delta+{\rm d}$. Differential forms on $\mathcal M$ are graded into a bi-complex by $\delta$ and ${\rm d}$:
\begin{equation*}
    \Omega^p(\mathcal M)=\bigoplus_{r=0}^p\Omega^{r,p-r}(\mathcal M),\ \Omega^{r,p-r}(\mathcal M)\xrightarrow{\ \delta\ }\Omega^{r,p-r+1}(\mathcal M),
\end{equation*}
with $\Omega^{r,p-r}(\mathcal M)=\emptyset$ for $r\geq2$ since there is only one $t$ variable. Importantly, each kinematic point $\{u\}$ locates a Riemann $t$-sphere in ${\cal M}$, and a $(1,1)$-form can be viewed as an $\Omega^1(\mathcal K)$-valued 1-form on the sphere. The line integral operator $\int_{a(u)}^{b(u)}:\Omega^1(S^2)\to\mathbb{C}$ can be extended to a linear map $\Omega^1(S^2)\otimes\Omega^1(\mathcal K)\to\Omega^1(\mathcal K)$, which defines an integration of $(1,1)$-forms.

To warm up, consider the total differential $\delta\mathcal T(u)$ of
\begin{equation}
    \mathcal T(u)=\int_{a(u)}^{b(u)}F(t;u)\,{\rm d}\log (t+c(u)),
\end{equation}
where $D F(t)$ is known~\footnote{From now on, we will omit the dependence on kinematics.}. Integrating by parts, $\delta\mathcal T$ has boundary contributions which are trivial to compute, as well as integral terms. A typical term $H(t)\,D\log(t+d)$ in $DF(t)$ contributes the integral term $\int_a^bH(t)\,\omega^{(1,1)}$, where $\omega^{(1,1)}$ is the $(1,1)$-component of the 2-form $\omega:=D\log (t+c) \wedge D\log (t+d)$. To obtain the symbol integration rule, we need to separate the $t$-dependence of $\omega^{(1,1)}$. This is done purely algebraically by matching residues, because $\omega^{(1,1)}\in\Omega^1(S^2)\otimes\Omega^1(\mathcal K)$ is a meromorphic 1-form on the $t$-sphere, which is determined by residues. Matching the residues at $t=-c$ and $t=-d$,
\begin{equation}
\omega^{(1,1)}={\rm d}\log\frac{t+c}{t+d}\wedge\delta\log(c-d).
\end{equation}
This way, we obtain the contribution to $\delta\mathcal T$:
\begin{equation}
\left(\int_a^bH(t)\,{\rm d}\log\frac{t+c}{t+d} \right)\delta\log(c-d).
\end{equation}
By definition, the above rule computes the $(W{-}1,1)$-coproduct of the weight-$W$ function $\mathcal T$, and iterating it yields the well-known symbol integration rule for linear symbol entries~\cite{Caron-Huot:2011dec,Li:2021bwg}.

Now we move to MPL symbol integrations involving square roots of quadratic polynomials, which usually requires rationalization and gets complicated when there are multiple square roots~\cite{Besier:2018jen}. We show that no explicit rationalization is needed from the 2-form perspective, and the method can be readily extended to elliptic cases. Our prototype is the integral
\begin{equation}
    \mathcal T=\int_a^b F(t)\,{\rm d}\log r(t),\quad DF(t)=H(t)\,D\log r_\Delta (t),
\end{equation}
where $r(t)=\frac{A(t)+\sqrt{R(t)}}{A(t)-\sqrt{R(t)}}$ and $r_\Delta(t)=\frac{B(t)+\sqrt{R(t)}\sqrt{\Delta(t)}}{B(t)-\sqrt{R(t)}\sqrt{\Delta(t)}}$. Here, $A(t), B(t)$ and even $R(t)$ can be arbitrary polynomials of $t$, but crucially $\Delta(t)$ is quadratic. Again, the key is to separate the $t$-dependence of $\omega:=D\log r(t) \wedge D\log r_\Delta(t)$. Note that it is parity-even under $\sqrt{R(t)} \to -\sqrt{R(t)}$. Hence, it is single-valued near $R(t)=0$ despite the apparent dependence on $\sqrt{R(t)}$, and the only branch points appear at $\Delta(t)=0$. Therefore, $\tilde\omega^{(1,1)}:=\sqrt{\Delta(t)}\,\omega^{(1,1)}$ is single-valued and meromorphic on the $t$-sphere. By matching residues of $\tilde\omega^{(1,1)}$, we obtain
\begin{equation}
    \omega^{(1,1)}=\sum_{t_0\in\{\text{poles of }\tilde\omega\}}\frac{\sqrt{\Delta(t_0)}\,{\rm d}t}{(t-t_0)\sqrt{\Delta(t)}}\wedge\mathop{\rm Res}_{t-t_0=0}\omega.
\end{equation}
We immediately obtain the integration rule in the same way as the linear-entry case:
\begin{equation}\label{eq:algrule}
    \begin{aligned}
        \delta\mathcal T=
        &\sum_{t_0\in\{\text{poles of }\omega\}}\left(\int_a^bH(t)\,\frac{\sqrt{\Delta(t_0)}\,{\rm d}t}{(t-t_0)\sqrt{\Delta(t)}}\right)\mathop{\rm Res}_{t-t_0=0}\omega\\
        &+F(b)\,\delta\log r(b)-F(a)\,\delta\log r(a),    \end{aligned}
\end{equation}
where the integration kernel can be nicely written as a ${\rm d}\log$ form since $\Delta(t)$ is quadratic, facilitating further iterations. A similar reasoning shows that the kernel becomes ${\rm d}\log(t-t_0)$ when there is no ``net'' square root $\sqrt{\Delta(t)}$ in the 2-form $\omega$.

Since square roots are carried along in our rules of symbol integration, no explicit rationalization (or any related subtleties~\cite{Li:2021bwg}) is involved. Moreover, our method generalizes existing ones and applies whenever the ``net'' square root $\sqrt{\Delta(t)}$ of $\omega$ has quadratic $\Delta(t)$.
The organization of results is also nicely suited for analyzing symbol structures of Feynman integrals --- the ``parity" of every square root is manifest, and a basis of independent last entries is obtained after only one iteration.

\begin{figure}[H]
    \centering
    \subfigure[]{\includegraphics[valign=c]{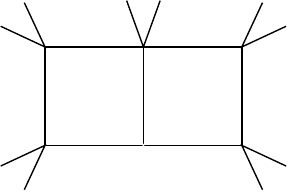}}
    \hspace{2em}\subfigure[]{\includegraphics[valign=c]{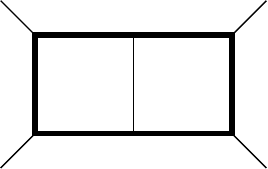}}
    \caption{Examples of MPL double-boxes, which depend on $5$ and $2$ kinematic variables, respectively.}
    \label{fig:MPLdb}
\end{figure}

We have applied \eqref{eq:algrule} to various non-trivial two-loop MPL integrals with square roots of quadratic polynomials. For example,the double-box with 5 massive legs (figure~\ref{fig:MPLdb}(a), $I_{17}$ in~\cite{He:2022ctv}) and the massless double-box with equally massive circumferential propagators (figure~\ref{fig:MPLdb}(b), $g_{10}$ in~\cite{Caron-Huot:2014lda}) has previously been computed only through canonical differential equations. Starting from the deformed hexagon representation with degenerations (see appendix~\ref{appA}), or the Mandelstam representation for the latter~\cite{Caron-Huot:2014lda}, we reproduce the 5 and 2 last entries and their symbols with very little work.


\subsection{Elliptic symbol integrations}
Next we consider elliptic integrals, where the prototype involves an elliptic curve $\mathcal E=\{(t,y)\,|\,y^2=P(t)\}\subseteq\mathbb{CP}^2$ and $P(t)$ is an irreducible cubic or quartic polynomial: 
\begin{equation}
    \mathcal I=\int_a^bF(t)\,\frac{{\rm d}t}{y(t)},\quad DF(t)=H(t)\,D\log r'_\Delta(t),
    \end{equation}
where $r'_\Delta(t):=\frac{B(t)+y(t)\sqrt{\Delta(t)}}{B(t)-y(t)\sqrt{\Delta(t)}}$ with $B(t)$ an arbitrary polynomial and $\Delta(t)$ quadratic. The key difference from MPL cases is that the integration kernel is no longer a ${\rm d}\log$ form, but we can still write it as a total differential $\frac{{\rm d}t}{y(t)}={\rm d}W(t)$ with$\quad W(t):=\int_\ast^t\frac{{\rm d}t'}{y(t')}$ for any reference point $\ast$. As in the MPL case we obtain \begin{equation}
    \delta\mathcal I=F(b)\,\delta W(b)-F(a)\,\delta W(a)+\int_a^bH(t)\,\omega^{(1,1)}.
\end{equation}

There are more than one possible choice for $W(t)$ even after we have fixed an initial point $\ast$, because the natural domain of the integrand is topologically a torus due to the branch points of $\sqrt{P(t)}$. The fundamental group is generated by two independent cycles $\gamma_{1,2}$, and we can freely add any multiples of $\gamma_{1,2}$ to the contour defining the integral $W(t)$, leading to definitions that differ by multiples of $\omega_{1,2}=\oint_{\gamma_{1,2}}{\rm d}t/y(t)$. Practically, after performing a bi-rational change of vaariables $(t,y)\mapsto(T,Y)$ to put $\mathcal E$ into Weierstrass form $Y^2=4T^3-g_2T-g_3$, we choose $W(t)=\wp^{-1}(T;g_2,g_3)$ for some branch of $\wp^{-1}$.
We also renormalize $\mathcal I$ and $W(t)$ with $\omega_1$: $\mathcal T:=\frac1{\omega_1}\mathcal I$, $w(t):=\frac1{\omega_1}W(t)$, and $\delta {\cal T}$ is of the same form as $\delta\mathcal I$ except $W(t) \to w(t)$.  

We wish to separate the $t$-dependence of the $(1,1)$-component of the 2-form $\omega:=Dw(t)\wedge D\log r'_\Delta(t)$ by matching residues. Since $\omega$ is parity-even under $y(t)\to-y(t)$, the ``net'' square root is $\sqrt{\Delta(t)}$, and defining $\tilde\omega^{(1,1)}=\sqrt{\Delta(t)}\,\omega^{(1,1)}$ eliminates the branch points at $\Delta(t)=0$. However, unlike MPL cases where $\tilde\omega^{(1,1)}$ is rational, the presence of $w(t)=\wp^{-1}(T)/\omega_1$ introduces extra branch cuts. Crossing the branch cuts leads to discontinuities in $\delta w(t)$ proportional to $\delta\tau$, where $\tau:=\omega_2/\omega_1$. Ultimately, the reason is that genus-1 curves have non-trivial moduli, and $\tau(u)$ depends on the kinematics.

It is not clear how to proceed directly, so we follow the proposal in~\cite{Morales:2022csr} and get around this problem by restricting to the subspace $\underline{\mathcal K}$ of $\mathcal K$ defined by $\delta\tau=0$. In other words, we focus on the elliptic symbol/coproduct up to period terms containing $\tau$.
Notationally, we use $\underline\delta$ to indicate the differential operator on $\underline{\mathcal K}$, and $\underline D={\rm d}+\underline\delta$. The restricted $(1,1)$-component   
\begin{equation}
\tilde\omega^{(1,1)}|_{\underline{\mathcal K}}=\sqrt{\Delta(t)}\left({\rm d}w(t)\wedge\delta\log r'_\Delta(t)+\underline\delta w(t)\wedge{\rm d}\log r'_\Delta(t)\right)
\end{equation}
is indeed an $\Omega^1(\underline{\mathcal K})$-valued meromorphic 1-form on the $t$-sphere. Incidentally, the restriction frees us from explicitly specifying the branch of $\wp^{-1}$ when defining $w(t)$.

We can now determine $\omega^{(1,1)}|_{\underline{\mathcal K}}$ by matching residues of $\tilde\omega^{(1,1)}|_{\underline{\mathcal K}}$,
and the residue computation is surprisingly easy: since ${\rm d}w(t)$ is holomorphic, the first term does not contribute at all, and all contributions come from singularities of $\log r'_\Delta(t)$. Denoting such singularities as $t_\pm$ which satisfy $B(t_\pm )\mp y(t_\pm)\sqrt{\Delta(t_\pm)}=0$, we have
\begin{equation}
    \omega^{(1,1)}|_{\underline{\mathcal K}}=\sum_{t_\pm}\pm \frac{\sqrt{\Delta(t_\pm)}\,{\rm d}t}{(t-t_\pm)\sqrt{\Delta(t)}}\wedge\underline\delta w(t_\pm)
\end{equation}
which gives the final result:
\begin{equation}\label{eq:ellrule}
    \begin{aligned}
        \underline\delta\mathcal T&=\sum_{t_\pm} \pm \left(\int_a^bH(t)\,\frac{\sqrt{\Delta(t_\pm)}\,{\rm d}t}{(t-t_\pm )\sqrt{\Delta(t)}}\right)\underline\delta w(t_\pm)
        \\&+F(b)\,\underline\delta w(b)-F(a)\,\underline\delta w(a).
    \end{aligned}
\end{equation}


\section{Application to double-$D$-gons in $D$ dimensions}
We now apply our method to conformal double-$D$-gons in $D$ dimensions, which, as reviewed in appendix~\ref{appA}, is expressed as a one-fold integral of deformed $2(D{-}1)$-gon with well-known symbol. We will obtain their last entries $\underline{\delta} w$ and the accompanying integrals by \eqref{eq:ellrule}. 

\subsection{Double-triangle integrals in $D=3$}
We start with the $D=3$ double-triangle, which can be represented as the integral of a deformed box \eqref{eq:dpoly}:
\begin{equation}
    \mathcal I_3=\int_0^\infty\frac{{\rm d}s}{\sqrt{-\mathcal Q(s^2)}}\,\langle\!\langle Q(s^2)\rangle\!\rangle,
\end{equation}
where we have performed a change of variable $t=s^2$ to get rid of the $\sqrt{t}$ in the denominator. The notation $\langle\!\langle Q(s^2)\rangle\!\rangle$ denotes the pure function \eqref{eq:quadpure} defined by the quadric $Q(s^2)$, and $\mathcal Q(s^2):=\det Q(s^2)$.

We first consider the special case with massless propagators, which depends on conformal cross-ratios $u=\frac{x_{12}^2x_{34}^2}{x_{13}^2x_{24}^2}$ and $v=\frac{x_{14}^2x_{23}^2}{x_{13}^2x_{24}^2}$ with $x_{ij}^2:=(x_i-x_j)^2$:
\begin{equation}
    \mathcal I_3=\includegraphics[scale=0.9,valign=c]{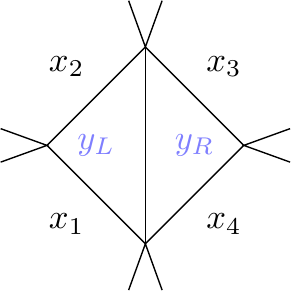}=\int_0^\infty{\rm d}s\includegraphics[scale=0.9,valign=c]{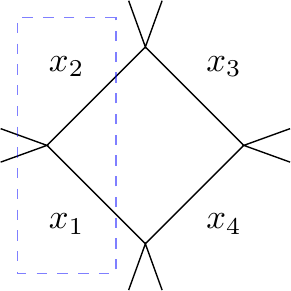}.
\end{equation}
The blue dashed box indicates deformation \eqref{eq:tdef}, and after some rescalings using projectivity, the quadric reads
\begin{equation}
    Q(s^2)=\left(\begin{matrix}
        0 & 1 & 1 & 1 \\
        1 & 0 & v & 1 \\
        1 & v & 0 & (1+s^2)u \\
        1 & 1 & (1+s^2)u & 0
    \end{matrix}\right).
\end{equation}
As usual, introduce $z,\bar z=\frac12(1+(1+s^2)u-v\pm\sqrt{\Delta(s)})$ and $\Delta(s)=(1-(1+s^2)u-v)^2-4(1+s^2)uv$ such that $(1+s^2)u=z\bar z$ and $v=(1-z)(1-\bar z)$. Then, $\langle\!\langle Q(s^2)\rangle\!\rangle$ is precisely the (deformed) four-mass box function~\cite{Drummond:2010cz}:
\begin{equation}
\langle\!\langle Q(s^2)\rangle\!\rangle=\log \frac{z\bar z}{v} \log \frac{1-z}{1-\bar{z}}-2{\rm Li}_2\left(\frac{-z}{1-z}\right)+2{\rm Li}_2\left(\frac{-\bar z}{1-\bar z}\right).
\end{equation}

The integral $\mathcal I_3$ is elliptic, involving the curve $y^2=-\mathcal Q(s^2)$. Define $\mathcal T_3:=\mathcal I_3/\omega_1$. It can be shown~\footnote{The bi-rational transformation $(s,y)\mapsto(S,Y)$ such that $Y^2=4S^3-g_2S-g_3$ maps $(0,\sqrt{-\mathcal Q(0)})$ to $Y=0$ and $(\infty,\infty)$ to $Y=\infty$. Since $(S,Y)=(\wp,\wp')$ is an isomorphism between the elliptic curve and the torus $\mathbb{C}/\langle1,\tau\rangle$, we see that $w(0)=\wp'^{-1}(0)=0$ mod $\langle1,\tau\rangle/2$ and $w(\infty)=\wp'^{-1}(\infty)=0$ mod $\langle1,\tau\rangle$.} that $w(0),w(\infty)=0$ mod $\langle1,\tau\rangle/2$, where $\langle1,\tau\rangle/2:=\frac12\mathbb{Z}+\frac12\mathbb{Z}\tau$ is the lattice generated by $1,\tau$ together with half lattice points. Therefore, $\underline\delta\mathcal T_3$ has no contribution from the boundary terms.
Note that the last entries of $\mathcal S(\langle\!\langle Q(s^2)\rangle\!\rangle)$ and the kernel $\frac{{\rm d}s}{\sqrt{-\mathcal Q(s^2)}}=i\frac{{\rm d}s}{\sqrt{\Delta(s)}}$
are both odd under $\sqrt{\Delta(s)}\to-\sqrt{\Delta(s)}$ so there are no ``net'' square roots. Applying the rules from the previous section,
\begin{equation}
    \underline\delta\mathcal T_3=\left(\int_0^\infty\log v\,{\rm d}\log(s-i)\right)\underline\delta w(i) + (i\to -i) 
\end{equation}
However, the elliptic curve $y^2=-\mathcal Q(s^2)$ is even under $s\to-s$, which implies $w(i)+w(-i)=0$ mod $\langle1,\tau\rangle$. Hence, there is only one independent last entry of $\underline\delta\mathcal T_3$:
\begin{equation}
    \underline\delta\mathcal T_3=\left(\int_0^\infty\log v\,{\rm d}\log\frac{s-i}{s+i}\right)\underline\delta w(i)=(i\pi\log v)\,\underline\delta w(i).
\end{equation}
In the last step, we have chosen to perform the integral on the function level, instead of using our symbol integration rules. Of course, the symbol integration rules still apply in this case, yielding a vanishing result because the symbol of $i\pi\log v$ as a weight-2 function is zero. The fact that $\underline\delta\mathcal T_3$ turns out to be proportional to $\pi$ is not unfamiliar for (MPL) integrals in three dimensions~\cite{Caron-Huot:2012sos,He:2022lfz,He:2023exb,Henn:2023pkc}.

\begin{figure}[H]
    \centering
    \subfigure[]{\includegraphics[scale=0.9,valign=c]{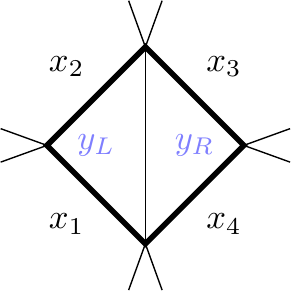}}\hspace{2em}\subfigure[]{\includegraphics[valign=c]{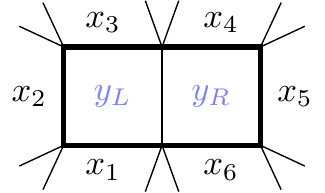}}
    \caption{Double-triangle and double-box with massive circumferential propagators, which depend on $6$ and $15$ cross-ratios.}
    \label{fig:dtdbM}
\end{figure}

The computation of the double-triangle with massive circumferential propagators (figure~\ref{fig:dtdbM}(a)) is entirely similar. Here, we merely record the result:
\begin{equation}
    \begin{aligned}
        \underline\delta\mathcal T_3^\text{massive}&=i\pi\log\frac{(1+u_{14})(1+u_{23})}{(1+u_{13})(1+u_{24})}\,\underline\delta w(i)\\
       &-i\pi\sum_{i=1}^4\log\frac{X_i+i\sqrt{R_i}}{X_i-i\sqrt{R_i}}\,\underline\delta w\left(\frac{\sqrt{R_i}}{\sqrt{U_i^2-1}}\right),
    \end{aligned}
\end{equation}
where   $u_{ij}=\frac{x_{ij}^2+m_i^2+m_j^2}{2m_im_j}$, and
\begin{align}
&    U_i=\begin{cases}
        u_{34},&i\in\{1,2\}\\
        u_{12},&i\in\{3,4\}
    \end{cases},
\quad X_i=1+\sum_{\substack{j,k\neq i\\j<k}}u_{jk},\\ 
&R_i=\mathcal Q^i_i(0)=1-\sum_{\substack{j,k\neq i\\j<k}}u_{jk}^2+2\prod_{\substack{j,k\neq i\\j<k}}u_{jk}.
\end{align}
Here, $\mathcal Q^i_i(0)$ is the minor of $Q(0)$ with the $i$-th row and column deleted. As a consistency check, $\underline\delta\mathcal T_3^\text{massive}$ has branch points at $u_{ij}=-1$ or $x_{ij}^2=-(m_i+m_j)^2$, exactly as predicted by Cutkosky's rules.

\subsection{Double-box integrals in $D=4$}

For the $D=4$ double-box (figure~\ref{fig:dtdbM}(b)), the starting point is the deformed hexagon (appendix \ref{appA}):
\begin{equation}
    \setlength\arraycolsep{1ex}
    \mathcal I=\int_0^\infty\frac{{\rm d}t}{\sqrt{-\mathcal Q(t)}}\,\langle\!\langle Q(t)\rangle\!\rangle,
\end{equation}
where $Q(t)$ is given by deforming the $6\times 6$ Gram matrix $G$ with 1 on the diagonal and $u_{ij}=\frac{x_{ij}^2+m_i^2+m_j^2}{2m_im_j}$ off the diagonal. Read off the symbol \eqref{eq:quadpure},
\begin{equation}
    \mathcal S(\langle\!\langle Q(t)\rangle\!\rangle)=\sum_{1\leq i<j\leq6}\mathcal S(\text{Box}_{ij}(t))\otimes\log R_{ij}(t),
\end{equation}
where $\text{Box}_{ij}(t)=\langle\!\langle Q^{ij}_{ij}(t)\rangle\!\rangle$ is obtained by deleting the $i$- and $j$-th row and column of $Q$, and the last entries $R_{ij}(t)=\underline{ij}$ are given by \eqref{eq:qentry}. Define the renormalized pure integral and last entries:
\begin{equation}
   \mathcal T=\frac1{\omega_1}\mathcal I,\quad w(t)=\frac1{\omega_1}\int^t\frac{{\rm d}t'}{\sqrt{-\mathcal Q(t')}}=\frac1{\omega_1}\wp^{-1}(T).
\end{equation}
Again, it can be shown that $w(\infty)=0$ mod $\langle1,\tau\rangle$, so there is no boundary term at $t=\infty$. The boundary term at $t=0$ is $-\mathcal S(\langle\!\langle Q(0)\rangle\!\rangle)\,\underline\delta w(0)$ representing the undeformed hexagon. For the integral terms, we need only consider singularities of $\log R_{ij}(t)$ located at $[\mathcal Q^i_j]^2=-\mathcal Q^{ij}_{ij}\mathcal Q \iff \mathcal Q^i_i\mathcal Q^j_j=0$, {\it i.e.}, zeros of $\mathcal Q^i_i$; here, $\mathcal Q^I_J:=\det Q^I_J(t)$ is the minor of $Q(t)$ with the rows (columns) labeled by $I$ ($J$) deleted.

Very nicely, the zeros of $5\times5$ minors $\mathcal Q^i_i$ are easy to obtain: for $i \in\{1,2,3\}$, the minor is quadratic in $t$ and has two roots $\{r_i^{(1)},r_i^{(2)}\}$; for $i\in\{4,5,6\}$, it is cubic with three roots $\{-1,r_i^{(1)},r_i^{(2)}\}$~\footnote{To show that $t=-1$ is a root, simply notice that when $t=-1$, the first three rows/columns are linearly dependent.}. Therefore, we have 13 different singularities coming from all possible $R_{ij}(t)$, which implies that the integral terms contributing to $\underline\delta\mathcal T$ have 13 possible last entries $\underline\delta w(t_0)$, where $t_0\in\{-1\}\cup\{r_i^{(1)},r_i^{(2)}\}_{i=1}^6$. In total,
\begin{equation}
    \underline\delta\mathcal T=-\mathcal S(\langle\!\langle Q(0)\rangle\!\rangle)\,\underline\delta w(0)+\sum_{t_0}\mathcal S(V_{t_0})\,\underline\delta w(t_0).
\end{equation}
which is what we expect: the kinematical space $\mathcal K$ is 15 dimensional, and one of the degrees of freedom is captured by the unknown $\delta\tau$ term, leaving 14 functionally independent last entries.

We can immediately write down an integral representation of the $(3,1)$-coproduct, as long as we keep track of the various signs:
\begin{align}
    V_{-1}&=\sum_{\substack{i\in\{1,2,3\}\\j\in\{4,5,6\}}}\int_0^\infty\,\text{Box}_{ij}(t)\,\frac{\sqrt{\Delta_{ij}(-1)}\,{\rm d}t}{(t+1)\sqrt{\Delta_{ij}(t)}},\\
    V_{r_i^{(a)}}&=\sum_{j\neq i}\pm \int_0^\infty\,\text{Box}_{ij}(t)\,\frac{\sqrt{\Delta_{ij}(r_i^{(a)})}\,{\rm d}t}{(t-r_i^{(a)})\sqrt{\Delta_{ij}(t)}},
\end{align}
where $\Delta_{ij}(t)$ is the box square root:
\begin{equation}
    \Delta_{ij}(t)=\begin{cases}
        \mathcal Q^{ij}_{ij}(t),&\text{if }\mathcal Q^{ij}_{ij}(-1)\neq0\\
        (t+1)^{-2}\mathcal Q^{ij}_{ij}(t),&\text{if }\mathcal Q^{ij}_{ij}(-1)=0
    \end{cases}.
\end{equation}
Nothing stops us from iterating our rules to obtain $\mathcal S(V_{t_0})$ explicitly, though the calculation is a bit tedious. We content ourselves with computing the symbol in the special case where all propagators are massless: the $12$ last entries $w(r_i^{(1,2)})$ satisfy linear relations and combine into $6$ independent ones (modulo $\langle1,\tau\rangle$). We have computed the $6$ accompanying weight-3 symbols and found perfect agreement with~\cite{Morales:2022csr}.


\subsection{Double-$D$-gon integrals in $D\geq 5$}
The $D{\leq}4$ and $D{\geq}5$ cases are different. The embedding space vectors live in $(D{+}2)$ dimensions, which implies that all $(D{+}3){\times}(D{+}3)$ minors of $G$ vanish for $D{\geq}5$ (no such minors exists for $D{\leq}4$). Therefore, up to the $(D{-}5)$-th derivatives vanish: $\mathcal Q(0){=}\mathcal Q'(0){=}\cdots{=}\mathcal Q^{(D{-}5)}(0){=}0$, which implies $\mathcal Q=-t^{D{-}4}P(t)$ where $P(0)\neq0$. Remarkably, for $D{\geq}5$, the integration kernel of $\mathcal I_D$ remains elliptic:
\begin{equation}
    \mathcal I_D=\int_0^\infty\frac{{\rm d}t}{\sqrt{P(t)}}\,F_{Q(t)},\quad\deg P(t)=3,\quad\forall D\geq5.
\end{equation}
Our method yields all the last entries of $\underline\delta\mathcal T_D$ together with the accompanying integrals for $\mathcal T_D=\mathcal I_D/\omega_1$.

The 2-form $\omega=\frac{Dt}{\sqrt{P(t)}}\wedge D\log\underline{\rho_2\rho_1}$ is proportional to $\sqrt{t^{D-4}\mathcal Q^{\rho_1\rho_2}_{\rho_1\rho_2}}=\sqrt{\Delta(t)}\times\text{rational}$, and after taking complete squares out of the square root, the ``net'' square root $\sqrt{\Delta(t)}$ is not necessarily quadratic. Hence, the kernel of the accompanying integral
\begin{equation}
    \int_0^\infty\langle\!\langle Q^{\rho_1\rho_2}_{\rho_1\rho_2}(t)\rangle\!\rangle\,\frac{\sqrt{\Delta(t_0)}\,{\rm d}t}{\sqrt{\Delta(t)}\,(t-t_0)}
\end{equation}
is not necessarily ${\rm d}\log$. Specifically, if $\Delta(t)$ is cubic or quartic, the accompanying integral itself is elliptic; and if $\deg\Delta(t)\geq5$, which first appears at $D{=}8$, the accompanying integral involves higher-genus curves and their symbology has not been studied in the literature. Our method provides partial results about these integrals, but conceivably we would miss even more terms because higher-genus curves have more periods.

\section{Conclusion and outlook}

We have proposed algebraic rules of (e)MPL symbol integration that efficiently computes the total differentials or $(W{-}1,1)$-coproducts of one-fold integrals of MPLs up to period terms, which can be iterated to produce the symbol. By exploiting the 2-form, we are able to sidestep rationalization completely, thus greatly improve on the existing method. We have checked our algorithm by reproducing (within minutes on a laptop using a very rough code) the results of some (e)MPL Feynman integrals, previously obtained through indirect methods.

Our algorithm applies nicely to the family of conformal double-$D$-gons in $D$ dimensions, possibly with massive circumferential propagators. In particular, we have computed the $(2,1)$-coproduct of the $D=3$ case on the function level, and have obtained an integral representation of the $(D{-}1,1)$-coproduct for $D\geq4$, up to period terms. Moreover, we have argued that unlike $D=3,4$ cases, the weight-$(D{-}1)$ integrals accompanying the last entries can involve elliptic and even higher-genus curves for large $D$. It would be extremely interesting to understand the symbol and the geometric interpretation of double-polygons, much like the well-known (one-loop) polygons~\cite{Spradlin:2011wp,Abreu:2017ptx, Arkani-Hamed:2017ahv,Herrmann:2019upk,Bourjaily:2019exo}.

Our method brings (elliptic) symbol integrations within reach for numerous other integrals. For example, it can be applied to integrals beyond double-triangles for higher-point two-loop amplitudes in ABJM theory~\cite{He:2022lfz}, and the recently studied family of elliptic ladder integrals~\cite{Cao:2023tpx,McLeod:2023qdf} can serve as an all-loop application of our method. Along this line, it would be highly desirable to systematize elliptic symbol integration to include different integration kernels~\cite{Broedel:2017kkb} and period terms. Another important question is how to extend our symbol integration rules to the function level, first for MPLs but eventually for eMPLs, now that we can avoid rationalization.

We expect that this computational method will reveal more structures of symbols and coproducts. The fact that symbol letters produced by our algorithm are closely related to singularities of the integrand may provide insight into the success of the recently proposed Schubert analysis~\cite{Yang:2022gko,He:2022ctv,He:2022tph,Morales:2022csr} in predicting (e)MPL symbol letters, and may further extend it to general spacetime dimensions. It would also be interesting to explore interpretations of the accompanying weight-$(W{-}1)$ integrals, along the lines of~\cite{Caron-Huot:2014lda} or~\cite{Herrmann:2019upk}, which is related to the diagrammatic coaction~\cite{Abreu:2017ptx,Abreu:2017enx,Abreu:2017mtm,Abreu:2019eyg,Abreu:2021vhb}. 

\begin{acknowledgments}
We thank Qu Cao, Zhenjie Li, Qinglin Yang and Chi Zhang for inspiring discussions and collaborations on related projects. The research of S. H. is supported in part by the National Natural Science Foundation of China under Grant No.11935013, 11947301, 12047502, 12047503. 
\end{acknowledgments}

\begin{appendix}
\section{The deformed polygon representation of double polygons}\label{appA}
In this appendix, we discuss the representation of double $D$-gons in $D$ dimensions as an integral of a deformed $2n$-gon~\cite{Cao:2023tpx} with $D=n+1\geq3$, where some of the $2n$ dual points may be identified. Schematically, we show that
\begin{equation}\label{eq:dpolysch}
    \mathcal I_D=\includegraphics[scale=0.9,valign=c]{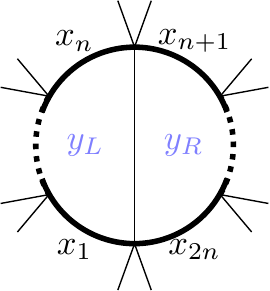}=\int_0^\infty t^{\frac{n-3}2}{\rm d}t\includegraphics[scale=0.9,valign=c]{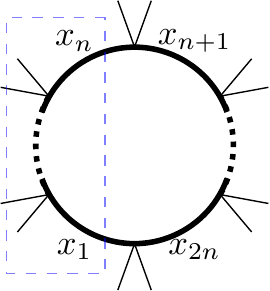}.
\end{equation}
The blue dashed box indicates $t$-deformation; see \eqref{eq:tdef}.

Consider the most general double polygon, with generically massive circumferential propagators,
\begin{equation}
    \mathcal I_D=\int{\rm d}^Dy_L{\rm d}^Dy_R\,D_{L,R}\prod_{i=1}^nD_{L,i}D_{R,n+i},
\end{equation}
where $D_{L,R}^{-1}=(y_L-y_R)^2$ and $D_{\ell,i}^{-1}=(y_\ell-x_i)^2+m_k^2$ for $\ell=L,R$. Using the embedding formalism and performing a loop-by-loop Feynman parametrization,
\begin{equation}
    \mathcal I_D=\int_0^\infty\frac{\langle\alpha{\rm d}^{2n-1}\alpha\rangle}{(R_1R_1)^{\frac{n-1}2}(R_2R_2)^{\frac{n+1}2}},
\end{equation}
where
\begin{equation}
    R_1=\alpha_1X_1+\cdots+\alpha_nX_n,\quad R_2=\alpha_1X_1+\cdots+\alpha_{2n}X_{2n},
\end{equation}
and the embedding space vectors $X_i^M{=}(x_i^\mu;x_i^2{+}m_i^2,1)$ have inner products $(X_iX_j){=}(x_i{-}x_j)^2{+}m_i^2{+}m_j^2$. Introducing a further Feynman parameter to combine the denominators,
\begin{equation}
    \mathcal I_D=\int_0^\infty t^{\frac{n-3}2}{\rm d}t\int_0^\infty\frac{\langle\alpha{\rm d}^{2n-1}\alpha\rangle}{(\alpha\cdot Q(t)\cdot\alpha)^n},
\end{equation}
where the quadric $\alpha\cdot Q(t)\cdot\alpha=t(R_1R_1)+(R_2R_2)$ represents a deformed $2n$-gon:
\vspace{1.2ex}
\begin{equation}\label{eq:tdef}
    \includegraphics[raise=-3.7em]{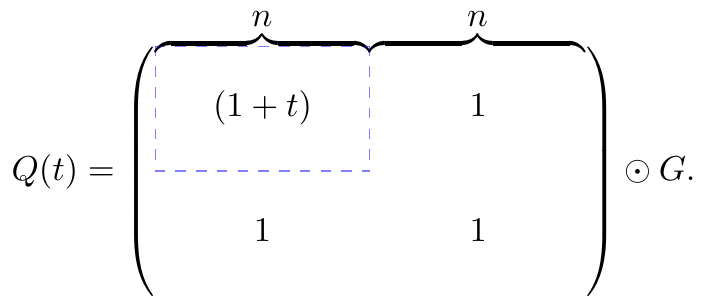}
\end{equation}
Here, the symbol ``$\odot$'' indicates element-wise multiplication, and the $(i,j)$-entry of the $2n\times2n$ Gram matrix $G$ is $(X_iX_j)$. We will often omit the $t$-dependence and denote $Q\equiv Q(t)$ and $G\equiv Q(0)$. Due to the projective nature of the quadric integral, we can freely rescale the $i$-th row and the $i$-th column by the same constant.

The result of the quadric integral is well-known~\cite{Arkani-Hamed:2017ahv}. It evaluates to an MPL function $\langle\!\langle Q\rangle\!\rangle/\sqrt{-\mathcal Q}$ with non-trivial leading singularity, where $\mathcal Q\equiv\det Q$ and $\langle\!\langle Q\rangle\!\rangle$ a pure function. In other words, we obtain the precise form of \eqref{eq:dpolysch}:
\begin{equation}\label{eq:dpoly}
    \mathcal I_D=\int_0^\infty\frac{t^{\frac{n-3}2}{\rm d}t}{\sqrt{-\mathcal Q}}\,\langle\!\langle Q(t)\rangle\!\rangle.
\end{equation}
The symbol of $\langle\!\langle Q(t)\rangle\!\rangle$ can be read off from the quadric:
\begin{equation}\label{eq:quadpure}
    \mathcal S(\langle\!\langle Q\rangle\!\rangle)=\sum_\rho\log\underline{\rho_{2n}\rho_{2n-1}}\otimes\cdots\otimes\log\underline{\rho_2\rho_1},
\end{equation}
where $\rho$ runs over all ordered partitions of $2n$ labels into $n$ symmetric pairs, and the symbol entries
\begin{equation}\label{eq:qentry}
    \underline{\rho_{2k}\rho_{2k-1}}=\frac{\mathcal Q^{\rho_{[2k-2]}\rho_{2k-1}}_{\rho_{[2k-2]}\rho_{2k}}+\sqrt{-\mathcal Q^{\rho_{[2k-2]}}_{\rho_{[2k-2]}}\mathcal Q^{\rho_{[2k]}}_{\rho_{[2k]}}}}{\mathcal Q^{\rho_{[2k-2]}\rho_{2k-1}}_{\rho_{[2k-2]}\rho_{2k}}-\sqrt{-\mathcal Q^{\rho_{[2k-2]}}_{\rho_{[2k-2]}}\mathcal Q^{\rho_{[2k]}}_{\rho_{[2k]}}}}.
\end{equation}
Here, $\rho_{[2k]}$ denotes the label set $\rho_1\cdots\rho_{2k}$ and $\mathcal Q^I_J=\det Q^I_J$ is the minor of $Q$ with the rows (columns) labeled by $I$ ($J$) deleted.
\end{appendix}

\bibliographystyle{apsrev4-1}
\bibliography{reference.bib}

\begin{thebibliography}{86}%
\makeatletter
\providecommand \@ifxundefined [1]{%
 \@ifx{#1\undefined}
}%
\providecommand \@ifnum [1]{%
 \ifnum #1\expandafter \@firstoftwo
 \else \expandafter \@secondoftwo
 \fi
}%
\providecommand \@ifx [1]{%
 \ifx #1\expandafter \@firstoftwo
 \else \expandafter \@secondoftwo
 \fi
}%
\providecommand \natexlab [1]{#1}%
\providecommand \enquote  [1]{``#1''}%
\providecommand \bibnamefont  [1]{#1}%
\providecommand \bibfnamefont [1]{#1}%
\providecommand \citenamefont [1]{#1}%
\providecommand \href@noop [0]{\@secondoftwo}%
\providecommand \href [0]{\begingroup \@sanitize@url \@href}%
\providecommand \@href[1]{\@@startlink{#1}\@@href}%
\providecommand \@@href[1]{\endgroup#1\@@endlink}%
\providecommand \@sanitize@url [0]{\catcode `\\12\catcode `\$12\catcode
  `\&12\catcode `\#12\catcode `\^12\catcode `\_12\catcode `\%12\relax}%
\providecommand \@@startlink[1]{}%
\providecommand \@@endlink[0]{}%
\providecommand \url  [0]{\begingroup\@sanitize@url \@url }%
\providecommand \@url [1]{\endgroup\@href {#1}{\urlprefix }}%
\providecommand \urlprefix  [0]{URL }%
\providecommand \Eprint [0]{\href }%
\providecommand \doibase [0]{http://dx.doi.org/}%
\providecommand \selectlanguage [0]{\@gobble}%
\providecommand \bibinfo  [0]{\@secondoftwo}%
\providecommand \bibfield  [0]{\@secondoftwo}%
\providecommand \translation [1]{[#1]}%
\providecommand \BibitemOpen [0]{}%
\providecommand \bibitemStop [0]{}%
\providecommand \bibitemNoStop [0]{.\EOS\space}%
\providecommand \EOS [0]{\spacefactor3000\relax}%
\providecommand \BibitemShut  [1]{\csname bibitem#1\endcsname}%
\let\auto@bib@innerbib\@empty
\bibitem [{\citenamefont {Chen}(1977)}]{Chen:1977oja}%
  \BibitemOpen
  \bibfield  {author} {\bibinfo {author} {\bibfnamefont {K.-T.}\ \bibnamefont
  {Chen}},\ }\href {\doibase 10.1090/S0002-9904-1977-14320-6} {\bibfield
  {journal} {\bibinfo  {journal} {Bull. Am. Math. Soc.}\ }\textbf {\bibinfo
  {volume} {83}},\ \bibinfo {pages} {831} (\bibinfo {year} {1977})}\BibitemShut
  {NoStop}%
\bibitem [{\citenamefont {Goncharov}(1995)}]{Goncharov1995GeometryOC}%
  \BibitemOpen
  \bibfield  {author} {\bibinfo {author} {\bibfnamefont {A.~B.}\ \bibnamefont
  {Goncharov}},\ }\href@noop {} {\bibfield  {journal} {\bibinfo  {journal}
  {Advances in Mathematics}\ }\textbf {\bibinfo {volume} {114}},\ \bibinfo
  {pages} {197} (\bibinfo {year} {1995})}\BibitemShut {NoStop}%
\bibitem [{\citenamefont {Goncharov}(1998)}]{Goncharov:1998kja}%
  \BibitemOpen
  \bibfield  {author} {\bibinfo {author} {\bibfnamefont {A.~B.}\ \bibnamefont
  {Goncharov}},\ }\href {\doibase 10.4310/MRL.1998.v5.n4.a7} {\bibfield
  {journal} {\bibinfo  {journal} {Math. Res. Lett.}\ }\textbf {\bibinfo
  {volume} {5}},\ \bibinfo {pages} {497} (\bibinfo {year} {1998})},\ \Eprint
  {http://arxiv.org/abs/1105.2076} {arXiv:1105.2076 [math.AG]} \BibitemShut
  {NoStop}%
\bibitem [{\citenamefont {Remiddi}\ and\ \citenamefont
  {Vermaseren}(2000)}]{Remiddi:1999ew}%
  \BibitemOpen
  \bibfield  {author} {\bibinfo {author} {\bibfnamefont {E.}~\bibnamefont
  {Remiddi}}\ and\ \bibinfo {author} {\bibfnamefont {J.~A.~M.}\ \bibnamefont
  {Vermaseren}},\ }\href {\doibase 10.1142/S0217751X00000367} {\bibfield
  {journal} {\bibinfo  {journal} {Int. J. Mod. Phys. A}\ }\textbf {\bibinfo
  {volume} {15}},\ \bibinfo {pages} {725} (\bibinfo {year} {2000})},\ \Eprint
  {http://arxiv.org/abs/hep-ph/9905237} {arXiv:hep-ph/9905237} \BibitemShut
  {NoStop}%
\bibitem [{\citenamefont {Borwein}\ \emph {et~al.}(2001)\citenamefont
  {Borwein}, \citenamefont {Bradley}, \citenamefont {Broadhurst},\ and\
  \citenamefont {Lisonek}}]{Borwein:1999js}%
  \BibitemOpen
  \bibfield  {author} {\bibinfo {author} {\bibfnamefont {J.~M.}\ \bibnamefont
  {Borwein}}, \bibinfo {author} {\bibfnamefont {D.~M.}\ \bibnamefont
  {Bradley}}, \bibinfo {author} {\bibfnamefont {D.~J.}\ \bibnamefont
  {Broadhurst}}, \ and\ \bibinfo {author} {\bibfnamefont {P.}~\bibnamefont
  {Lisonek}},\ }\href {\doibase 10.1090/S0002-9947-00-02616-7} {\bibfield
  {journal} {\bibinfo  {journal} {Trans. Am. Math. Soc.}\ }\textbf {\bibinfo
  {volume} {353}},\ \bibinfo {pages} {907} (\bibinfo {year} {2001})},\ \Eprint
  {http://arxiv.org/abs/math/9910045} {arXiv:math/9910045} \BibitemShut
  {NoStop}%
\bibitem [{\citenamefont {Moch}\ \emph {et~al.}(2002)\citenamefont {Moch},
  \citenamefont {Uwer},\ and\ \citenamefont {Weinzierl}}]{Moch:2001zr}%
  \BibitemOpen
  \bibfield  {author} {\bibinfo {author} {\bibfnamefont {S.}~\bibnamefont
  {Moch}}, \bibinfo {author} {\bibfnamefont {P.}~\bibnamefont {Uwer}}, \ and\
  \bibinfo {author} {\bibfnamefont {S.}~\bibnamefont {Weinzierl}},\ }\href
  {\doibase 10.1063/1.1471366} {\bibfield  {journal} {\bibinfo  {journal} {J.
  Math. Phys.}\ }\textbf {\bibinfo {volume} {43}},\ \bibinfo {pages} {3363}
  (\bibinfo {year} {2002})},\ \Eprint {http://arxiv.org/abs/hep-ph/0110083}
  {arXiv:hep-ph/0110083} \BibitemShut {NoStop}%
\bibitem [{\citenamefont {Kotikov}(1991{\natexlab{a}})}]{Kotikov:1990kg}%
  \BibitemOpen
  \bibfield  {author} {\bibinfo {author} {\bibfnamefont {A.~V.}\ \bibnamefont
  {Kotikov}},\ }\href {\doibase 10.1016/0370-2693(91)90413-K} {\bibfield
  {journal} {\bibinfo  {journal} {Phys. Lett. B}\ }\textbf {\bibinfo {volume}
  {254}},\ \bibinfo {pages} {158} (\bibinfo {year}
  {1991}{\natexlab{a}})}\BibitemShut {NoStop}%
\bibitem [{\citenamefont {Kotikov}(1991{\natexlab{b}})}]{Kotikov:1991pm}%
  \BibitemOpen
  \bibfield  {author} {\bibinfo {author} {\bibfnamefont {A.~V.}\ \bibnamefont
  {Kotikov}},\ }\href {\doibase 10.1016/0370-2693(91)90536-Y} {\bibfield
  {journal} {\bibinfo  {journal} {Phys. Lett. B}\ }\textbf {\bibinfo {volume}
  {267}},\ \bibinfo {pages} {123} (\bibinfo {year} {1991}{\natexlab{b}})},\
  \bibinfo {note} {[Erratum: Phys.Lett.B 295, 409--409 (1992)]}\BibitemShut
  {NoStop}%
\bibitem [{\citenamefont {Remiddi}(1997)}]{Remiddi:1997ny}%
  \BibitemOpen
  \bibfield  {author} {\bibinfo {author} {\bibfnamefont {E.}~\bibnamefont
  {Remiddi}},\ }\href {\doibase 10.1007/BF03185566} {\bibfield  {journal}
  {\bibinfo  {journal} {Nuovo Cim. A}\ }\textbf {\bibinfo {volume} {110}},\
  \bibinfo {pages} {1435} (\bibinfo {year} {1997})},\ \Eprint
  {http://arxiv.org/abs/hep-th/9711188} {arXiv:hep-th/9711188} \BibitemShut
  {NoStop}%
\bibitem [{\citenamefont {Gehrmann}\ and\ \citenamefont
  {Remiddi}(2000)}]{Gehrmann:1999as}%
  \BibitemOpen
  \bibfield  {author} {\bibinfo {author} {\bibfnamefont {T.}~\bibnamefont
  {Gehrmann}}\ and\ \bibinfo {author} {\bibfnamefont {E.}~\bibnamefont
  {Remiddi}},\ }\href {\doibase 10.1016/S0550-3213(00)00223-6} {\bibfield
  {journal} {\bibinfo  {journal} {Nucl. Phys. B}\ }\textbf {\bibinfo {volume}
  {580}},\ \bibinfo {pages} {485} (\bibinfo {year} {2000})},\ \Eprint
  {http://arxiv.org/abs/hep-ph/9912329} {arXiv:hep-ph/9912329} \BibitemShut
  {NoStop}%
\bibitem [{\citenamefont {Henn}(2013)}]{Henn:2013pwa}%
  \BibitemOpen
  \bibfield  {author} {\bibinfo {author} {\bibfnamefont {J.~M.}\ \bibnamefont
  {Henn}},\ }\href {\doibase 10.1103/PhysRevLett.110.251601} {\bibfield
  {journal} {\bibinfo  {journal} {Phys. Rev. Lett.}\ }\textbf {\bibinfo
  {volume} {110}},\ \bibinfo {pages} {251601} (\bibinfo {year} {2013})},\
  \Eprint {http://arxiv.org/abs/1304.1806} {arXiv:1304.1806 [hep-th]}
  \BibitemShut {NoStop}%
\bibitem [{\citenamefont {Bourjaily}\ \emph {et~al.}(2018)\citenamefont
  {Bourjaily}, \citenamefont {McLeod}, \citenamefont {von Hippel},\ and\
  \citenamefont {Wilhelm}}]{Bourjaily:2018aeq}%
  \BibitemOpen
  \bibfield  {author} {\bibinfo {author} {\bibfnamefont {J.~L.}\ \bibnamefont
  {Bourjaily}}, \bibinfo {author} {\bibfnamefont {A.~J.}\ \bibnamefont
  {McLeod}}, \bibinfo {author} {\bibfnamefont {M.}~\bibnamefont {von Hippel}},
  \ and\ \bibinfo {author} {\bibfnamefont {M.}~\bibnamefont {Wilhelm}},\ }\href
  {\doibase 10.1007/JHEP08(2018)184} {\bibfield  {journal} {\bibinfo  {journal}
  {JHEP}\ }\textbf {\bibinfo {volume} {08}},\ \bibinfo {pages} {184} (\bibinfo
  {year} {2018})},\ \Eprint {http://arxiv.org/abs/1805.10281} {arXiv:1805.10281
  [hep-th]} \BibitemShut {NoStop}%
\bibitem [{\citenamefont {Bourjaily}\ \emph {et~al.}(2019)\citenamefont
  {Bourjaily}, \citenamefont {Dulat},\ and\ \citenamefont
  {Panzer}}]{Bourjaily:2019jrk}%
  \BibitemOpen
  \bibfield  {author} {\bibinfo {author} {\bibfnamefont {J.~L.}\ \bibnamefont
  {Bourjaily}}, \bibinfo {author} {\bibfnamefont {F.}~\bibnamefont {Dulat}}, \
  and\ \bibinfo {author} {\bibfnamefont {E.}~\bibnamefont {Panzer}},\ }\href
  {\doibase 10.1016/j.nuclphysb.2019.03.022} {\bibfield  {journal} {\bibinfo
  {journal} {Nucl. Phys. B}\ }\textbf {\bibinfo {volume} {942}},\ \bibinfo
  {pages} {251} (\bibinfo {year} {2019})},\ \Eprint
  {http://arxiv.org/abs/1901.02887} {arXiv:1901.02887 [hep-th]} \BibitemShut
  {NoStop}%
\bibitem [{\citenamefont {Bourjaily}\ \emph
  {et~al.}(2020{\natexlab{a}})\citenamefont {Bourjaily}, \citenamefont {Volk},\
  and\ \citenamefont {Von~Hippel}}]{Bourjaily:2019vby}%
  \BibitemOpen
  \bibfield  {author} {\bibinfo {author} {\bibfnamefont {J.~L.}\ \bibnamefont
  {Bourjaily}}, \bibinfo {author} {\bibfnamefont {M.}~\bibnamefont {Volk}}, \
  and\ \bibinfo {author} {\bibfnamefont {M.}~\bibnamefont {Von~Hippel}},\
  }\href {\doibase 10.1007/JHEP02(2020)095} {\bibfield  {journal} {\bibinfo
  {journal} {JHEP}\ }\textbf {\bibinfo {volume} {02}},\ \bibinfo {pages} {095}
  (\bibinfo {year} {2020}{\natexlab{a}})},\ \Eprint
  {http://arxiv.org/abs/1912.05690} {arXiv:1912.05690 [hep-th]} \BibitemShut
  {NoStop}%
\bibitem [{\citenamefont {Bourjaily}\ \emph {et~al.}(2021)\citenamefont
  {Bourjaily}, \citenamefont {He}, \citenamefont {McLeod}, \citenamefont
  {Spradlin}, \citenamefont {Vergu}, \citenamefont {Volk}, \citenamefont {von
  Hippel},\ and\ \citenamefont {Wilhelm}}]{Bourjaily:2021lnz}%
  \BibitemOpen
  \bibfield  {author} {\bibinfo {author} {\bibfnamefont {J.~L.}\ \bibnamefont
  {Bourjaily}}, \bibinfo {author} {\bibfnamefont {Y.-H.}\ \bibnamefont {He}},
  \bibinfo {author} {\bibfnamefont {A.~J.}\ \bibnamefont {McLeod}}, \bibinfo
  {author} {\bibfnamefont {M.}~\bibnamefont {Spradlin}}, \bibinfo {author}
  {\bibfnamefont {C.}~\bibnamefont {Vergu}}, \bibinfo {author} {\bibfnamefont
  {M.}~\bibnamefont {Volk}}, \bibinfo {author} {\bibfnamefont {M.}~\bibnamefont
  {von Hippel}}, \ and\ \bibinfo {author} {\bibfnamefont {M.}~\bibnamefont
  {Wilhelm}},\ }in\ \href {\doibase 10.1007/978-3-030-80219-6_5} {\emph
  {\bibinfo {booktitle} {{Antidifferentiation and the Calculation of Feynman
  Amplitudes}}}}\ (\bibinfo {year} {2021})\ \Eprint
  {http://arxiv.org/abs/2103.15423} {arXiv:2103.15423 [hep-th]} \BibitemShut
  {NoStop}%
\bibitem [{\citenamefont {Panzer}(2015)}]{Panzer:2014caa}%
  \BibitemOpen
  \bibfield  {author} {\bibinfo {author} {\bibfnamefont {E.}~\bibnamefont
  {Panzer}},\ }\href {\doibase 10.1016/j.cpc.2014.10.019} {\bibfield  {journal}
  {\bibinfo  {journal} {Comput. Phys. Commun.}\ }\textbf {\bibinfo {volume}
  {188}},\ \bibinfo {pages} {148} (\bibinfo {year} {2015})},\ \Eprint
  {http://arxiv.org/abs/1403.3385} {arXiv:1403.3385 [hep-th]} \BibitemShut
  {NoStop}%
\bibitem [{\citenamefont {Duhr}\ and\ \citenamefont
  {Dulat}(2019)}]{Duhr:2019tlz}%
  \BibitemOpen
  \bibfield  {author} {\bibinfo {author} {\bibfnamefont {C.}~\bibnamefont
  {Duhr}}\ and\ \bibinfo {author} {\bibfnamefont {F.}~\bibnamefont {Dulat}},\
  }\href {\doibase 10.1007/JHEP08(2019)135} {\bibfield  {journal} {\bibinfo
  {journal} {JHEP}\ }\textbf {\bibinfo {volume} {08}},\ \bibinfo {pages} {135}
  (\bibinfo {year} {2019})},\ \Eprint {http://arxiv.org/abs/1904.07279}
  {arXiv:1904.07279 [hep-th]} \BibitemShut {NoStop}%
\bibitem [{\citenamefont {Li}(2021)}]{Gint}%
  \BibitemOpen
  \bibfield  {author} {\bibinfo {author} {\bibfnamefont {Z.}~\bibnamefont
  {Li}},\ }\href@noop {} {\enquote {\bibinfo {title}
  {Multiple-polylogarithm},}\ }\bibinfo {howpublished}
  {\url{https://github.com/munuxi/Multiple-Polylogarithm}} (\bibinfo {year}
  {2021})\BibitemShut {NoStop}%
\bibitem [{\citenamefont {Caron-Huot}(2011)}]{Caron-Huot:2011zgw}%
  \BibitemOpen
  \bibfield  {author} {\bibinfo {author} {\bibfnamefont {S.}~\bibnamefont
  {Caron-Huot}},\ }\href {\doibase 10.1007/JHEP12(2011)066} {\bibfield
  {journal} {\bibinfo  {journal} {JHEP}\ }\textbf {\bibinfo {volume} {12}},\
  \bibinfo {pages} {066} (\bibinfo {year} {2011})},\ \Eprint
  {http://arxiv.org/abs/1105.5606} {arXiv:1105.5606 [hep-th]} \BibitemShut
  {NoStop}%
\bibitem [{\citenamefont {He}\ \emph {et~al.}(2021{\natexlab{a}})\citenamefont
  {He}, \citenamefont {Li}, \citenamefont {Tang},\ and\ \citenamefont
  {Yang}}]{He:2020uxy}%
  \BibitemOpen
  \bibfield  {author} {\bibinfo {author} {\bibfnamefont {S.}~\bibnamefont
  {He}}, \bibinfo {author} {\bibfnamefont {Z.}~\bibnamefont {Li}}, \bibinfo
  {author} {\bibfnamefont {Y.}~\bibnamefont {Tang}}, \ and\ \bibinfo {author}
  {\bibfnamefont {Q.}~\bibnamefont {Yang}},\ }\href {\doibase
  10.1007/JHEP05(2021)052} {\bibfield  {journal} {\bibinfo  {journal} {JHEP}\
  }\textbf {\bibinfo {volume} {05}},\ \bibinfo {pages} {052} (\bibinfo {year}
  {2021}{\natexlab{a}})},\ \Eprint {http://arxiv.org/abs/2012.13094}
  {arXiv:2012.13094 [hep-th]} \BibitemShut {NoStop}%
\bibitem [{\citenamefont {He}\ \emph {et~al.}(2021{\natexlab{b}})\citenamefont
  {He}, \citenamefont {Li}, \citenamefont {Yang},\ and\ \citenamefont
  {Zhang}}]{He:2020lcu}%
  \BibitemOpen
  \bibfield  {author} {\bibinfo {author} {\bibfnamefont {S.}~\bibnamefont
  {He}}, \bibinfo {author} {\bibfnamefont {Z.}~\bibnamefont {Li}}, \bibinfo
  {author} {\bibfnamefont {Q.}~\bibnamefont {Yang}}, \ and\ \bibinfo {author}
  {\bibfnamefont {C.}~\bibnamefont {Zhang}},\ }\href {\doibase
  10.1103/PhysRevLett.126.231601} {\bibfield  {journal} {\bibinfo  {journal}
  {Phys. Rev. Lett.}\ }\textbf {\bibinfo {volume} {126}},\ \bibinfo {pages}
  {231601} (\bibinfo {year} {2021}{\natexlab{b}})},\ \Eprint
  {http://arxiv.org/abs/2012.15042} {arXiv:2012.15042 [hep-th]} \BibitemShut
  {NoStop}%
\bibitem [{\citenamefont {Chicherin}\ \emph {et~al.}(2018)\citenamefont
  {Chicherin}, \citenamefont {Henn},\ and\ \citenamefont
  {Mitev}}]{Chicherin:2017dob}%
  \BibitemOpen
  \bibfield  {author} {\bibinfo {author} {\bibfnamefont {D.}~\bibnamefont
  {Chicherin}}, \bibinfo {author} {\bibfnamefont {J.}~\bibnamefont {Henn}}, \
  and\ \bibinfo {author} {\bibfnamefont {V.}~\bibnamefont {Mitev}},\ }\href
  {\doibase 10.1007/JHEP05(2018)164} {\bibfield  {journal} {\bibinfo  {journal}
  {JHEP}\ }\textbf {\bibinfo {volume} {05}},\ \bibinfo {pages} {164} (\bibinfo
  {year} {2018})},\ \Eprint {http://arxiv.org/abs/1712.09610} {arXiv:1712.09610
  [hep-th]} \BibitemShut {NoStop}%
\bibitem [{\citenamefont {Henn}\ \emph {et~al.}(2018)\citenamefont {Henn},
  \citenamefont {Herrmann},\ and\ \citenamefont
  {Parra-Martinez}}]{Henn:2018cdp}%
  \BibitemOpen
  \bibfield  {author} {\bibinfo {author} {\bibfnamefont {J.}~\bibnamefont
  {Henn}}, \bibinfo {author} {\bibfnamefont {E.}~\bibnamefont {Herrmann}}, \
  and\ \bibinfo {author} {\bibfnamefont {J.}~\bibnamefont {Parra-Martinez}},\
  }\href {\doibase 10.1007/JHEP10(2018)059} {\bibfield  {journal} {\bibinfo
  {journal} {JHEP}\ }\textbf {\bibinfo {volume} {10}},\ \bibinfo {pages} {059}
  (\bibinfo {year} {2018})},\ \Eprint {http://arxiv.org/abs/1806.06072}
  {arXiv:1806.06072 [hep-th]} \BibitemShut {NoStop}%
\bibitem [{\citenamefont {He}\ \emph {et~al.}(2021{\natexlab{c}})\citenamefont
  {He}, \citenamefont {Li},\ and\ \citenamefont {Yang}}]{He:2021esx}%
  \BibitemOpen
  \bibfield  {author} {\bibinfo {author} {\bibfnamefont {S.}~\bibnamefont
  {He}}, \bibinfo {author} {\bibfnamefont {Z.}~\bibnamefont {Li}}, \ and\
  \bibinfo {author} {\bibfnamefont {Q.}~\bibnamefont {Yang}},\ }\href {\doibase
  10.1007/JHEP06(2021)119} {\bibfield  {journal} {\bibinfo  {journal} {JHEP}\
  }\textbf {\bibinfo {volume} {06}},\ \bibinfo {pages} {119} (\bibinfo {year}
  {2021}{\natexlab{c}})},\ \Eprint {http://arxiv.org/abs/2103.02796}
  {arXiv:2103.02796 [hep-th]} \BibitemShut {NoStop}%
\bibitem [{\citenamefont {He}\ \emph {et~al.}(2021{\natexlab{d}})\citenamefont
  {He}, \citenamefont {Li},\ and\ \citenamefont {Yang}}]{He:2021eec}%
  \BibitemOpen
  \bibfield  {author} {\bibinfo {author} {\bibfnamefont {S.}~\bibnamefont
  {He}}, \bibinfo {author} {\bibfnamefont {Z.}~\bibnamefont {Li}}, \ and\
  \bibinfo {author} {\bibfnamefont {Q.}~\bibnamefont {Yang}},\ }\href@noop {}
  {\  (\bibinfo {year} {2021}{\natexlab{d}})},\ \Eprint
  {http://arxiv.org/abs/2112.11842} {arXiv:2112.11842 [hep-th]} \BibitemShut
  {NoStop}%
\bibitem [{\citenamefont {He}\ \emph {et~al.}(2022{\natexlab{a}})\citenamefont
  {He}, \citenamefont {Li}, \citenamefont {Ma}, \citenamefont {Wu},
  \citenamefont {Yang},\ and\ \citenamefont {Zhang}}]{He:2022ctv}%
  \BibitemOpen
  \bibfield  {author} {\bibinfo {author} {\bibfnamefont {S.}~\bibnamefont
  {He}}, \bibinfo {author} {\bibfnamefont {Z.}~\bibnamefont {Li}}, \bibinfo
  {author} {\bibfnamefont {R.}~\bibnamefont {Ma}}, \bibinfo {author}
  {\bibfnamefont {Z.}~\bibnamefont {Wu}}, \bibinfo {author} {\bibfnamefont
  {Q.}~\bibnamefont {Yang}}, \ and\ \bibinfo {author} {\bibfnamefont
  {Y.}~\bibnamefont {Zhang}},\ }\href {\doibase 10.1007/JHEP10(2022)165}
  {\bibfield  {journal} {\bibinfo  {journal} {JHEP}\ }\textbf {\bibinfo
  {volume} {10}},\ \bibinfo {pages} {165} (\bibinfo {year}
  {2022}{\natexlab{a}})},\ \Eprint {http://arxiv.org/abs/2206.04609}
  {arXiv:2206.04609 [hep-th]} \BibitemShut {NoStop}%
\bibitem [{\citenamefont {Goncharov}(2005)}]{Goncharov:2005sla}%
  \BibitemOpen
  \bibfield  {author} {\bibinfo {author} {\bibfnamefont {A.~B.}\ \bibnamefont
  {Goncharov}},\ }\href {\doibase 10.1215/S0012-7094-04-12822-2} {\bibfield
  {journal} {\bibinfo  {journal} {Duke Math. J.}\ }\textbf {\bibinfo {volume}
  {128}},\ \bibinfo {pages} {209} (\bibinfo {year} {2005})},\ \Eprint
  {http://arxiv.org/abs/math/0208144} {arXiv:math/0208144} \BibitemShut
  {NoStop}%
\bibitem [{\citenamefont {Goncharov}\ \emph {et~al.}(2010)\citenamefont
  {Goncharov}, \citenamefont {Spradlin}, \citenamefont {Vergu},\ and\
  \citenamefont {Volovich}}]{Goncharov:2010jf}%
  \BibitemOpen
  \bibfield  {author} {\bibinfo {author} {\bibfnamefont {A.~B.}\ \bibnamefont
  {Goncharov}}, \bibinfo {author} {\bibfnamefont {M.}~\bibnamefont {Spradlin}},
  \bibinfo {author} {\bibfnamefont {C.}~\bibnamefont {Vergu}}, \ and\ \bibinfo
  {author} {\bibfnamefont {A.}~\bibnamefont {Volovich}},\ }\href {\doibase
  10.1103/PhysRevLett.105.151605} {\bibfield  {journal} {\bibinfo  {journal}
  {Phys. Rev. Lett.}\ }\textbf {\bibinfo {volume} {105}},\ \bibinfo {pages}
  {151605} (\bibinfo {year} {2010})},\ \Eprint {http://arxiv.org/abs/1006.5703}
  {arXiv:1006.5703 [hep-th]} \BibitemShut {NoStop}%
\bibitem [{\citenamefont {Spradlin}\ and\ \citenamefont
  {Volovich}(2011)}]{Spradlin:2011wp}%
  \BibitemOpen
  \bibfield  {author} {\bibinfo {author} {\bibfnamefont {M.}~\bibnamefont
  {Spradlin}}\ and\ \bibinfo {author} {\bibfnamefont {A.}~\bibnamefont
  {Volovich}},\ }\href {\doibase 10.1007/JHEP11(2011)084} {\bibfield  {journal}
  {\bibinfo  {journal} {JHEP}\ }\textbf {\bibinfo {volume} {11}},\ \bibinfo
  {pages} {084} (\bibinfo {year} {2011})},\ \Eprint
  {http://arxiv.org/abs/1105.2024} {arXiv:1105.2024 [hep-th]} \BibitemShut
  {NoStop}%
\bibitem [{\citenamefont {Duhr}\ \emph {et~al.}(2012)\citenamefont {Duhr},
  \citenamefont {Gangl},\ and\ \citenamefont {Rhodes}}]{Duhr:2011zq}%
  \BibitemOpen
  \bibfield  {author} {\bibinfo {author} {\bibfnamefont {C.}~\bibnamefont
  {Duhr}}, \bibinfo {author} {\bibfnamefont {H.}~\bibnamefont {Gangl}}, \ and\
  \bibinfo {author} {\bibfnamefont {J.~R.}\ \bibnamefont {Rhodes}},\ }\href
  {\doibase 10.1007/JHEP10(2012)075} {\bibfield  {journal} {\bibinfo  {journal}
  {JHEP}\ }\textbf {\bibinfo {volume} {10}},\ \bibinfo {pages} {075} (\bibinfo
  {year} {2012})},\ \Eprint {http://arxiv.org/abs/1110.0458} {arXiv:1110.0458
  [math-ph]} \BibitemShut {NoStop}%
\bibitem [{\citenamefont {Duhr}(2012)}]{Duhr:2012fh}%
  \BibitemOpen
  \bibfield  {author} {\bibinfo {author} {\bibfnamefont {C.}~\bibnamefont
  {Duhr}},\ }\href {\doibase 10.1007/JHEP08(2012)043} {\bibfield  {journal}
  {\bibinfo  {journal} {JHEP}\ }\textbf {\bibinfo {volume} {08}},\ \bibinfo
  {pages} {043} (\bibinfo {year} {2012})},\ \Eprint
  {http://arxiv.org/abs/1203.0454} {arXiv:1203.0454 [hep-ph]} \BibitemShut
  {NoStop}%
\bibitem [{\citenamefont {Bourjaily}\ \emph {et~al.}(2022)\citenamefont
  {Bourjaily} \emph {et~al.}}]{Bourjaily:2022bwx}%
  \BibitemOpen
  \bibfield  {author} {\bibinfo {author} {\bibfnamefont {J.~L.}\ \bibnamefont
  {Bourjaily}} \emph {et~al.},\ }in\ \href@noop {} {\emph {\bibinfo {booktitle}
  {{2022 Snowmass Summer Study}}}}\ (\bibinfo {year} {2022})\ \Eprint
  {http://arxiv.org/abs/2203.07088} {arXiv:2203.07088 [hep-ph]} \BibitemShut
  {NoStop}%
\bibitem [{\citenamefont {Laporta}\ and\ \citenamefont
  {Remiddi}(2005)}]{Laporta:2004rb}%
  \BibitemOpen
  \bibfield  {author} {\bibinfo {author} {\bibfnamefont {S.}~\bibnamefont
  {Laporta}}\ and\ \bibinfo {author} {\bibfnamefont {E.}~\bibnamefont
  {Remiddi}},\ }\href {\doibase 10.1016/j.nuclphysb.2004.10.044} {\bibfield
  {journal} {\bibinfo  {journal} {Nucl. Phys. B}\ }\textbf {\bibinfo {volume}
  {704}},\ \bibinfo {pages} {349} (\bibinfo {year} {2005})},\ \Eprint
  {http://arxiv.org/abs/hep-ph/0406160} {arXiv:hep-ph/0406160} \BibitemShut
  {NoStop}%
\bibitem [{\citenamefont {Brown}\ and\ \citenamefont
  {Levin}(2011)}]{brown2011multiple}%
  \BibitemOpen
  \bibfield  {author} {\bibinfo {author} {\bibfnamefont {F.}~\bibnamefont
  {Brown}}\ and\ \bibinfo {author} {\bibfnamefont {A.}~\bibnamefont {Levin}},\
  }\href@noop {} {\  (\bibinfo {year} {2011})},\ \Eprint
  {http://arxiv.org/abs/1110.6917} {arXiv:1110.6917 [math]} \BibitemShut
  {NoStop}%
\bibitem [{\citenamefont {Muller-Stach}\ \emph {et~al.}(2012)\citenamefont
  {Muller-Stach}, \citenamefont {Weinzierl},\ and\ \citenamefont
  {Zayadeh}}]{Muller-Stach:2012tgz}%
  \BibitemOpen
  \bibfield  {author} {\bibinfo {author} {\bibfnamefont {S.}~\bibnamefont
  {Muller-Stach}}, \bibinfo {author} {\bibfnamefont {S.}~\bibnamefont
  {Weinzierl}}, \ and\ \bibinfo {author} {\bibfnamefont {R.}~\bibnamefont
  {Zayadeh}},\ }\href {\doibase 10.22323/1.151.0005} {\bibfield  {journal}
  {\bibinfo  {journal} {PoS}\ }\textbf {\bibinfo {volume} {LL2012}},\ \bibinfo
  {pages} {005} (\bibinfo {year} {2012})},\ \Eprint
  {http://arxiv.org/abs/1209.3714} {arXiv:1209.3714 [hep-ph]} \BibitemShut
  {NoStop}%
\bibitem [{\citenamefont {Adams}\ \emph {et~al.}(2013)\citenamefont {Adams},
  \citenamefont {Bogner},\ and\ \citenamefont {Weinzierl}}]{Adams:2013nia}%
  \BibitemOpen
  \bibfield  {author} {\bibinfo {author} {\bibfnamefont {L.}~\bibnamefont
  {Adams}}, \bibinfo {author} {\bibfnamefont {C.}~\bibnamefont {Bogner}}, \
  and\ \bibinfo {author} {\bibfnamefont {S.}~\bibnamefont {Weinzierl}},\ }\href
  {\doibase 10.1063/1.4804996} {\bibfield  {journal} {\bibinfo  {journal} {J.
  Math. Phys.}\ }\textbf {\bibinfo {volume} {54}},\ \bibinfo {pages} {052303}
  (\bibinfo {year} {2013})},\ \Eprint {http://arxiv.org/abs/1302.7004}
  {arXiv:1302.7004 [hep-ph]} \BibitemShut {NoStop}%
\bibitem [{\citenamefont {Bloch}\ and\ \citenamefont
  {Vanhove}(2015)}]{Bloch:2013tra}%
  \BibitemOpen
  \bibfield  {author} {\bibinfo {author} {\bibfnamefont {S.}~\bibnamefont
  {Bloch}}\ and\ \bibinfo {author} {\bibfnamefont {P.}~\bibnamefont
  {Vanhove}},\ }\href {\doibase 10.1016/j.jnt.2014.09.032} {\bibfield
  {journal} {\bibinfo  {journal} {J. Number Theor.}\ }\textbf {\bibinfo
  {volume} {148}},\ \bibinfo {pages} {328} (\bibinfo {year} {2015})},\ \Eprint
  {http://arxiv.org/abs/1309.5865} {arXiv:1309.5865 [hep-th]} \BibitemShut
  {NoStop}%
\bibitem [{\citenamefont {Adams}\ \emph {et~al.}(2014)\citenamefont {Adams},
  \citenamefont {Bogner},\ and\ \citenamefont {Weinzierl}}]{Adams:2014vja}%
  \BibitemOpen
  \bibfield  {author} {\bibinfo {author} {\bibfnamefont {L.}~\bibnamefont
  {Adams}}, \bibinfo {author} {\bibfnamefont {C.}~\bibnamefont {Bogner}}, \
  and\ \bibinfo {author} {\bibfnamefont {S.}~\bibnamefont {Weinzierl}},\ }\href
  {\doibase 10.1063/1.4896563} {\bibfield  {journal} {\bibinfo  {journal} {J.
  Math. Phys.}\ }\textbf {\bibinfo {volume} {55}},\ \bibinfo {pages} {102301}
  (\bibinfo {year} {2014})},\ \Eprint {http://arxiv.org/abs/1405.5640}
  {arXiv:1405.5640 [hep-ph]} \BibitemShut {NoStop}%
\bibitem [{\citenamefont {Adams}\ \emph {et~al.}(2015)\citenamefont {Adams},
  \citenamefont {Bogner},\ and\ \citenamefont {Weinzierl}}]{Adams:2015gva}%
  \BibitemOpen
  \bibfield  {author} {\bibinfo {author} {\bibfnamefont {L.}~\bibnamefont
  {Adams}}, \bibinfo {author} {\bibfnamefont {C.}~\bibnamefont {Bogner}}, \
  and\ \bibinfo {author} {\bibfnamefont {S.}~\bibnamefont {Weinzierl}},\ }\href
  {\doibase 10.1063/1.4926985} {\bibfield  {journal} {\bibinfo  {journal} {J.
  Math. Phys.}\ }\textbf {\bibinfo {volume} {56}},\ \bibinfo {pages} {072303}
  (\bibinfo {year} {2015})},\ \Eprint {http://arxiv.org/abs/1504.03255}
  {arXiv:1504.03255 [hep-ph]} \BibitemShut {NoStop}%
\bibitem [{\citenamefont {Adams}\ \emph
  {et~al.}(2016{\natexlab{a}})\citenamefont {Adams}, \citenamefont {Bogner},\
  and\ \citenamefont {Weinzierl}}]{Adams:2015ydq}%
  \BibitemOpen
  \bibfield  {author} {\bibinfo {author} {\bibfnamefont {L.}~\bibnamefont
  {Adams}}, \bibinfo {author} {\bibfnamefont {C.}~\bibnamefont {Bogner}}, \
  and\ \bibinfo {author} {\bibfnamefont {S.}~\bibnamefont {Weinzierl}},\ }\href
  {\doibase 10.1063/1.4944722} {\bibfield  {journal} {\bibinfo  {journal} {J.
  Math. Phys.}\ }\textbf {\bibinfo {volume} {57}},\ \bibinfo {pages} {032304}
  (\bibinfo {year} {2016}{\natexlab{a}})},\ \Eprint
  {http://arxiv.org/abs/1512.05630} {arXiv:1512.05630 [hep-ph]} \BibitemShut
  {NoStop}%
\bibitem [{\citenamefont {Adams}\ \emph
  {et~al.}(2016{\natexlab{b}})\citenamefont {Adams}, \citenamefont {Bogner},
  \citenamefont {Schweitzer},\ and\ \citenamefont {Weinzierl}}]{Adams:2016xah}%
  \BibitemOpen
  \bibfield  {author} {\bibinfo {author} {\bibfnamefont {L.}~\bibnamefont
  {Adams}}, \bibinfo {author} {\bibfnamefont {C.}~\bibnamefont {Bogner}},
  \bibinfo {author} {\bibfnamefont {A.}~\bibnamefont {Schweitzer}}, \ and\
  \bibinfo {author} {\bibfnamefont {S.}~\bibnamefont {Weinzierl}},\ }\href
  {\doibase 10.1063/1.4969060} {\bibfield  {journal} {\bibinfo  {journal} {J.
  Math. Phys.}\ }\textbf {\bibinfo {volume} {57}},\ \bibinfo {pages} {122302}
  (\bibinfo {year} {2016}{\natexlab{b}})},\ \Eprint
  {http://arxiv.org/abs/1607.01571} {arXiv:1607.01571 [hep-ph]} \BibitemShut
  {NoStop}%
\bibitem [{\citenamefont {Adams}\ \emph {et~al.}(2017)\citenamefont {Adams},
  \citenamefont {Chaubey},\ and\ \citenamefont {Weinzierl}}]{Adams:2017tga}%
  \BibitemOpen
  \bibfield  {author} {\bibinfo {author} {\bibfnamefont {L.}~\bibnamefont
  {Adams}}, \bibinfo {author} {\bibfnamefont {E.}~\bibnamefont {Chaubey}}, \
  and\ \bibinfo {author} {\bibfnamefont {S.}~\bibnamefont {Weinzierl}},\ }\href
  {\doibase 10.1103/PhysRevLett.118.141602} {\bibfield  {journal} {\bibinfo
  {journal} {Phys. Rev. Lett.}\ }\textbf {\bibinfo {volume} {118}},\ \bibinfo
  {pages} {141602} (\bibinfo {year} {2017})},\ \Eprint
  {http://arxiv.org/abs/1702.04279} {arXiv:1702.04279 [hep-ph]} \BibitemShut
  {NoStop}%
\bibitem [{\citenamefont {Adams}\ and\ \citenamefont
  {Weinzierl}(2018{\natexlab{a}})}]{Adams:2017ejb}%
  \BibitemOpen
  \bibfield  {author} {\bibinfo {author} {\bibfnamefont {L.}~\bibnamefont
  {Adams}}\ and\ \bibinfo {author} {\bibfnamefont {S.}~\bibnamefont
  {Weinzierl}},\ }\href {\doibase 10.4310/CNTP.2018.v12.n2.a1} {\bibfield
  {journal} {\bibinfo  {journal} {Commun. Num. Theor. Phys.}\ }\textbf
  {\bibinfo {volume} {12}},\ \bibinfo {pages} {193} (\bibinfo {year}
  {2018}{\natexlab{a}})},\ \Eprint {http://arxiv.org/abs/1704.08895}
  {arXiv:1704.08895 [hep-ph]} \BibitemShut {NoStop}%
\bibitem [{\citenamefont {Bogner}\ \emph {et~al.}(2017)\citenamefont {Bogner},
  \citenamefont {Schweitzer},\ and\ \citenamefont
  {Weinzierl}}]{Bogner:2017vim}%
  \BibitemOpen
  \bibfield  {author} {\bibinfo {author} {\bibfnamefont {C.}~\bibnamefont
  {Bogner}}, \bibinfo {author} {\bibfnamefont {A.}~\bibnamefont {Schweitzer}},
  \ and\ \bibinfo {author} {\bibfnamefont {S.}~\bibnamefont {Weinzierl}},\
  }\href {\doibase 10.1016/j.nuclphysb.2017.07.008} {\bibfield  {journal}
  {\bibinfo  {journal} {Nucl. Phys. B}\ }\textbf {\bibinfo {volume} {922}},\
  \bibinfo {pages} {528} (\bibinfo {year} {2017})},\ \Eprint
  {http://arxiv.org/abs/1705.08952} {arXiv:1705.08952 [hep-ph]} \BibitemShut
  {NoStop}%
\bibitem [{\citenamefont {Broedel}\ \emph
  {et~al.}(2018{\natexlab{a}})\citenamefont {Broedel}, \citenamefont {Duhr},
  \citenamefont {Dulat},\ and\ \citenamefont {Tancredi}}]{Broedel:2017kkb}%
  \BibitemOpen
  \bibfield  {author} {\bibinfo {author} {\bibfnamefont {J.}~\bibnamefont
  {Broedel}}, \bibinfo {author} {\bibfnamefont {C.}~\bibnamefont {Duhr}},
  \bibinfo {author} {\bibfnamefont {F.}~\bibnamefont {Dulat}}, \ and\ \bibinfo
  {author} {\bibfnamefont {L.}~\bibnamefont {Tancredi}},\ }\href {\doibase
  10.1007/JHEP05(2018)093} {\bibfield  {journal} {\bibinfo  {journal} {JHEP}\
  }\textbf {\bibinfo {volume} {05}},\ \bibinfo {pages} {093} (\bibinfo {year}
  {2018}{\natexlab{a}})},\ \Eprint {http://arxiv.org/abs/1712.07089}
  {arXiv:1712.07089 [hep-th]} \BibitemShut {NoStop}%
\bibitem [{\citenamefont {Broedel}\ \emph
  {et~al.}(2018{\natexlab{b}})\citenamefont {Broedel}, \citenamefont {Duhr},
  \citenamefont {Dulat},\ and\ \citenamefont {Tancredi}}]{Broedel:2017siw}%
  \BibitemOpen
  \bibfield  {author} {\bibinfo {author} {\bibfnamefont {J.}~\bibnamefont
  {Broedel}}, \bibinfo {author} {\bibfnamefont {C.}~\bibnamefont {Duhr}},
  \bibinfo {author} {\bibfnamefont {F.}~\bibnamefont {Dulat}}, \ and\ \bibinfo
  {author} {\bibfnamefont {L.}~\bibnamefont {Tancredi}},\ }\href {\doibase
  10.1103/PhysRevD.97.116009} {\bibfield  {journal} {\bibinfo  {journal} {Phys.
  Rev. D}\ }\textbf {\bibinfo {volume} {97}},\ \bibinfo {pages} {116009}
  (\bibinfo {year} {2018}{\natexlab{b}})},\ \Eprint
  {http://arxiv.org/abs/1712.07095} {arXiv:1712.07095 [hep-ph]} \BibitemShut
  {NoStop}%
\bibitem [{\citenamefont {Adams}\ and\ \citenamefont
  {Weinzierl}(2018{\natexlab{b}})}]{Adams:2018yfj}%
  \BibitemOpen
  \bibfield  {author} {\bibinfo {author} {\bibfnamefont {L.}~\bibnamefont
  {Adams}}\ and\ \bibinfo {author} {\bibfnamefont {S.}~\bibnamefont
  {Weinzierl}},\ }\href {\doibase 10.1016/j.physletb.2018.04.002} {\bibfield
  {journal} {\bibinfo  {journal} {Phys. Lett. B}\ }\textbf {\bibinfo {volume}
  {781}},\ \bibinfo {pages} {270} (\bibinfo {year} {2018}{\natexlab{b}})},\
  \Eprint {http://arxiv.org/abs/1802.05020} {arXiv:1802.05020 [hep-ph]}
  \BibitemShut {NoStop}%
\bibitem [{\citenamefont {Broedel}\ \emph
  {et~al.}(2018{\natexlab{c}})\citenamefont {Broedel}, \citenamefont {Duhr},
  \citenamefont {Dulat}, \citenamefont {Penante},\ and\ \citenamefont
  {Tancredi}}]{Broedel:2018iwv}%
  \BibitemOpen
  \bibfield  {author} {\bibinfo {author} {\bibfnamefont {J.}~\bibnamefont
  {Broedel}}, \bibinfo {author} {\bibfnamefont {C.}~\bibnamefont {Duhr}},
  \bibinfo {author} {\bibfnamefont {F.}~\bibnamefont {Dulat}}, \bibinfo
  {author} {\bibfnamefont {B.}~\bibnamefont {Penante}}, \ and\ \bibinfo
  {author} {\bibfnamefont {L.}~\bibnamefont {Tancredi}},\ }\href {\doibase
  10.1007/JHEP08(2018)014} {\bibfield  {journal} {\bibinfo  {journal} {JHEP}\
  }\textbf {\bibinfo {volume} {08}},\ \bibinfo {pages} {014} (\bibinfo {year}
  {2018}{\natexlab{c}})},\ \Eprint {http://arxiv.org/abs/1803.10256}
  {arXiv:1803.10256 [hep-th]} \BibitemShut {NoStop}%
\bibitem [{\citenamefont {Broedel}\ \emph
  {et~al.}(2019{\natexlab{a}})\citenamefont {Broedel}, \citenamefont {Duhr},
  \citenamefont {Dulat}, \citenamefont {Penante},\ and\ \citenamefont
  {Tancredi}}]{Broedel:2018qkq}%
  \BibitemOpen
  \bibfield  {author} {\bibinfo {author} {\bibfnamefont {J.}~\bibnamefont
  {Broedel}}, \bibinfo {author} {\bibfnamefont {C.}~\bibnamefont {Duhr}},
  \bibinfo {author} {\bibfnamefont {F.}~\bibnamefont {Dulat}}, \bibinfo
  {author} {\bibfnamefont {B.}~\bibnamefont {Penante}}, \ and\ \bibinfo
  {author} {\bibfnamefont {L.}~\bibnamefont {Tancredi}},\ }\href {\doibase
  10.1007/JHEP01(2019)023} {\bibfield  {journal} {\bibinfo  {journal} {JHEP}\
  }\textbf {\bibinfo {volume} {01}},\ \bibinfo {pages} {023} (\bibinfo {year}
  {2019}{\natexlab{a}})},\ \Eprint {http://arxiv.org/abs/1809.10698}
  {arXiv:1809.10698 [hep-th]} \BibitemShut {NoStop}%
\bibitem [{\citenamefont {H\"onemann}\ \emph {et~al.}(2018)\citenamefont
  {H\"onemann}, \citenamefont {Tempest},\ and\ \citenamefont
  {Weinzierl}}]{Honemann:2018mrb}%
  \BibitemOpen
  \bibfield  {author} {\bibinfo {author} {\bibfnamefont {I.}~\bibnamefont
  {H\"onemann}}, \bibinfo {author} {\bibfnamefont {K.}~\bibnamefont {Tempest}},
  \ and\ \bibinfo {author} {\bibfnamefont {S.}~\bibnamefont {Weinzierl}},\
  }\href {\doibase 10.1103/PhysRevD.98.113008} {\bibfield  {journal} {\bibinfo
  {journal} {Phys. Rev. D}\ }\textbf {\bibinfo {volume} {98}},\ \bibinfo
  {pages} {113008} (\bibinfo {year} {2018})},\ \Eprint
  {http://arxiv.org/abs/1811.09308} {arXiv:1811.09308 [hep-ph]} \BibitemShut
  {NoStop}%
\bibitem [{\citenamefont {Broedel}\ \emph
  {et~al.}(2019{\natexlab{b}})\citenamefont {Broedel}, \citenamefont {Duhr},
  \citenamefont {Dulat}, \citenamefont {Penante},\ and\ \citenamefont
  {Tancredi}}]{Broedel:2019hyg}%
  \BibitemOpen
  \bibfield  {author} {\bibinfo {author} {\bibfnamefont {J.}~\bibnamefont
  {Broedel}}, \bibinfo {author} {\bibfnamefont {C.}~\bibnamefont {Duhr}},
  \bibinfo {author} {\bibfnamefont {F.}~\bibnamefont {Dulat}}, \bibinfo
  {author} {\bibfnamefont {B.}~\bibnamefont {Penante}}, \ and\ \bibinfo
  {author} {\bibfnamefont {L.}~\bibnamefont {Tancredi}},\ }\href {\doibase
  10.1007/JHEP05(2019)120} {\bibfield  {journal} {\bibinfo  {journal} {JHEP}\
  }\textbf {\bibinfo {volume} {05}},\ \bibinfo {pages} {120} (\bibinfo {year}
  {2019}{\natexlab{b}})},\ \Eprint {http://arxiv.org/abs/1902.09971}
  {arXiv:1902.09971 [hep-ph]} \BibitemShut {NoStop}%
\bibitem [{\citenamefont {Bogner}\ \emph {et~al.}(2020)\citenamefont {Bogner},
  \citenamefont {M\"uller-Stach},\ and\ \citenamefont
  {Weinzierl}}]{Bogner:2019lfa}%
  \BibitemOpen
  \bibfield  {author} {\bibinfo {author} {\bibfnamefont {C.}~\bibnamefont
  {Bogner}}, \bibinfo {author} {\bibfnamefont {S.}~\bibnamefont
  {M\"uller-Stach}}, \ and\ \bibinfo {author} {\bibfnamefont {S.}~\bibnamefont
  {Weinzierl}},\ }\href {\doibase 10.1016/j.nuclphysb.2020.114991} {\bibfield
  {journal} {\bibinfo  {journal} {Nucl. Phys. B}\ }\textbf {\bibinfo {volume}
  {954}},\ \bibinfo {pages} {114991} (\bibinfo {year} {2020})},\ \Eprint
  {http://arxiv.org/abs/1907.01251} {arXiv:1907.01251 [hep-th]} \BibitemShut
  {NoStop}%
\bibitem [{\citenamefont {Duhr}\ and\ \citenamefont
  {Tancredi}(2020)}]{Duhr:2019rrs}%
  \BibitemOpen
  \bibfield  {author} {\bibinfo {author} {\bibfnamefont {C.}~\bibnamefont
  {Duhr}}\ and\ \bibinfo {author} {\bibfnamefont {L.}~\bibnamefont
  {Tancredi}},\ }\href {\doibase 10.1007/JHEP02(2020)105} {\bibfield  {journal}
  {\bibinfo  {journal} {JHEP}\ }\textbf {\bibinfo {volume} {02}},\ \bibinfo
  {pages} {105} (\bibinfo {year} {2020})},\ \Eprint
  {http://arxiv.org/abs/1912.00077} {arXiv:1912.00077 [hep-th]} \BibitemShut
  {NoStop}%
\bibitem [{\citenamefont {Walden}\ and\ \citenamefont
  {Weinzierl}(2021)}]{Walden:2020odh}%
  \BibitemOpen
  \bibfield  {author} {\bibinfo {author} {\bibfnamefont {M.}~\bibnamefont
  {Walden}}\ and\ \bibinfo {author} {\bibfnamefont {S.}~\bibnamefont
  {Weinzierl}},\ }\href {\doibase 10.1016/j.cpc.2021.108020} {\bibfield
  {journal} {\bibinfo  {journal} {Comput. Phys. Commun.}\ }\textbf {\bibinfo
  {volume} {265}},\ \bibinfo {pages} {108020} (\bibinfo {year} {2021})},\
  \Eprint {http://arxiv.org/abs/2010.05271} {arXiv:2010.05271 [hep-ph]}
  \BibitemShut {NoStop}%
\bibitem [{\citenamefont {Weinzierl}(2021)}]{Weinzierl:2020fyx}%
  \BibitemOpen
  \bibfield  {author} {\bibinfo {author} {\bibfnamefont {S.}~\bibnamefont
  {Weinzierl}},\ }\href {\doibase 10.1016/j.nuclphysb.2021.115309} {\bibfield
  {journal} {\bibinfo  {journal} {Nucl. Phys. B}\ }\textbf {\bibinfo {volume}
  {964}},\ \bibinfo {pages} {115309} (\bibinfo {year} {2021})},\ \Eprint
  {http://arxiv.org/abs/2011.07311} {arXiv:2011.07311 [hep-th]} \BibitemShut
  {NoStop}%
\bibitem [{\citenamefont {Kristensson}\ \emph {et~al.}(2021)\citenamefont
  {Kristensson}, \citenamefont {Wilhelm},\ and\ \citenamefont
  {Zhang}}]{Kristensson:2021ani}%
  \BibitemOpen
  \bibfield  {author} {\bibinfo {author} {\bibfnamefont {A.}~\bibnamefont
  {Kristensson}}, \bibinfo {author} {\bibfnamefont {M.}~\bibnamefont
  {Wilhelm}}, \ and\ \bibinfo {author} {\bibfnamefont {C.}~\bibnamefont
  {Zhang}},\ }\href {\doibase 10.1103/PhysRevLett.127.251603} {\bibfield
  {journal} {\bibinfo  {journal} {Phys. Rev. Lett.}\ }\textbf {\bibinfo
  {volume} {127}},\ \bibinfo {pages} {251603} (\bibinfo {year} {2021})},\
  \Eprint {http://arxiv.org/abs/2106.14902} {arXiv:2106.14902 [hep-th]}
  \BibitemShut {NoStop}%
\bibitem [{\citenamefont {Wilhelm}\ and\ \citenamefont
  {Zhang}(2023)}]{Wilhelm:2022wow}%
  \BibitemOpen
  \bibfield  {author} {\bibinfo {author} {\bibfnamefont {M.}~\bibnamefont
  {Wilhelm}}\ and\ \bibinfo {author} {\bibfnamefont {C.}~\bibnamefont
  {Zhang}},\ }\href {\doibase 10.1007/JHEP01(2023)089} {\bibfield  {journal}
  {\bibinfo  {journal} {JHEP}\ }\textbf {\bibinfo {volume} {01}},\ \bibinfo
  {pages} {089} (\bibinfo {year} {2023})},\ \Eprint
  {http://arxiv.org/abs/2206.08378} {arXiv:2206.08378 [hep-th]} \BibitemShut
  {NoStop}%
\bibitem [{\citenamefont {Giroux}\ and\ \citenamefont
  {Pokraka}(2022)}]{Giroux:2022wav}%
  \BibitemOpen
  \bibfield  {author} {\bibinfo {author} {\bibfnamefont {M.}~\bibnamefont
  {Giroux}}\ and\ \bibinfo {author} {\bibfnamefont {A.}~\bibnamefont
  {Pokraka}},\ }\href@noop {} {\  (\bibinfo {year} {2022})},\ \Eprint
  {http://arxiv.org/abs/2210.09898} {arXiv:2210.09898 [hep-th]} \BibitemShut
  {NoStop}%
\bibitem [{\citenamefont {Morales}\ \emph {et~al.}(2022)\citenamefont
  {Morales}, \citenamefont {Spiering}, \citenamefont {Wilhelm}, \citenamefont
  {Yang},\ and\ \citenamefont {Zhang}}]{Morales:2022csr}%
  \BibitemOpen
  \bibfield  {author} {\bibinfo {author} {\bibfnamefont {R.}~\bibnamefont
  {Morales}}, \bibinfo {author} {\bibfnamefont {A.}~\bibnamefont {Spiering}},
  \bibinfo {author} {\bibfnamefont {M.}~\bibnamefont {Wilhelm}}, \bibinfo
  {author} {\bibfnamefont {Q.}~\bibnamefont {Yang}}, \ and\ \bibinfo {author}
  {\bibfnamefont {C.}~\bibnamefont {Zhang}},\ }\href@noop {} {\  (\bibinfo
  {year} {2022})},\ \Eprint {http://arxiv.org/abs/2212.09762} {arXiv:2212.09762
  [hep-th]} \BibitemShut {NoStop}%
\bibitem [{Note1()}]{Note1}%
  \BibitemOpen
  \bibinfo {note} {We mainly consider finite integrals in integer dimensions,
  but the method applies to each order in $\epsilon $ to dimensionally
  regularized integrals, and to integrals with mass regulators. It can also be
  used for the direct integration of amplitudes, Wilson loops~\cite
  {Caron-Huot:2011dec}, \protect \emph {etc.}}\BibitemShut {Stop}%
\bibitem [{\citenamefont {Caron-Huot}\ and\ \citenamefont
  {He}(2012)}]{Caron-Huot:2011dec}%
  \BibitemOpen
  \bibfield  {author} {\bibinfo {author} {\bibfnamefont {S.}~\bibnamefont
  {Caron-Huot}}\ and\ \bibinfo {author} {\bibfnamefont {S.}~\bibnamefont
  {He}},\ }\href {\doibase 10.1007/JHEP07(2012)174} {\bibfield  {journal}
  {\bibinfo  {journal} {JHEP}\ }\textbf {\bibinfo {volume} {07}},\ \bibinfo
  {pages} {174} (\bibinfo {year} {2012})},\ \Eprint
  {http://arxiv.org/abs/1112.1060} {arXiv:1112.1060 [hep-th]} \BibitemShut
  {NoStop}%
\bibitem [{\citenamefont {Paulos}\ \emph {et~al.}(2012)\citenamefont {Paulos},
  \citenamefont {Spradlin},\ and\ \citenamefont {Volovich}}]{Paulos:2012nu}%
  \BibitemOpen
  \bibfield  {author} {\bibinfo {author} {\bibfnamefont {M.~F.}\ \bibnamefont
  {Paulos}}, \bibinfo {author} {\bibfnamefont {M.}~\bibnamefont {Spradlin}}, \
  and\ \bibinfo {author} {\bibfnamefont {A.}~\bibnamefont {Volovich}},\ }\href
  {\doibase 10.1007/JHEP08(2012)072} {\bibfield  {journal} {\bibinfo  {journal}
  {JHEP}\ }\textbf {\bibinfo {volume} {08}},\ \bibinfo {pages} {072} (\bibinfo
  {year} {2012})},\ \Eprint {http://arxiv.org/abs/1203.6362} {arXiv:1203.6362
  [hep-th]} \BibitemShut {NoStop}%
\bibitem [{\citenamefont {Nandan}\ \emph {et~al.}(2013)\citenamefont {Nandan},
  \citenamefont {Paulos}, \citenamefont {Spradlin},\ and\ \citenamefont
  {Volovich}}]{Nandan:2013ip}%
  \BibitemOpen
  \bibfield  {author} {\bibinfo {author} {\bibfnamefont {D.}~\bibnamefont
  {Nandan}}, \bibinfo {author} {\bibfnamefont {M.~F.}\ \bibnamefont {Paulos}},
  \bibinfo {author} {\bibfnamefont {M.}~\bibnamefont {Spradlin}}, \ and\
  \bibinfo {author} {\bibfnamefont {A.}~\bibnamefont {Volovich}},\ }\href
  {\doibase 10.1007/JHEP05(2013)105} {\bibfield  {journal} {\bibinfo  {journal}
  {JHEP}\ }\textbf {\bibinfo {volume} {05}},\ \bibinfo {pages} {105} (\bibinfo
  {year} {2013})},\ \Eprint {http://arxiv.org/abs/1301.2500} {arXiv:1301.2500
  [hep-th]} \BibitemShut {NoStop}%
\bibitem [{\citenamefont {Abreu}\ \emph
  {et~al.}(2017{\natexlab{a}})\citenamefont {Abreu}, \citenamefont {Britto},
  \citenamefont {Duhr},\ and\ \citenamefont {Gardi}}]{Abreu:2017ptx}%
  \BibitemOpen
  \bibfield  {author} {\bibinfo {author} {\bibfnamefont {S.}~\bibnamefont
  {Abreu}}, \bibinfo {author} {\bibfnamefont {R.}~\bibnamefont {Britto}},
  \bibinfo {author} {\bibfnamefont {C.}~\bibnamefont {Duhr}}, \ and\ \bibinfo
  {author} {\bibfnamefont {E.}~\bibnamefont {Gardi}},\ }\href {\doibase
  10.1007/JHEP06(2017)114} {\bibfield  {journal} {\bibinfo  {journal} {JHEP}\
  }\textbf {\bibinfo {volume} {06}},\ \bibinfo {pages} {114} (\bibinfo {year}
  {2017}{\natexlab{a}})},\ \Eprint {http://arxiv.org/abs/1702.03163}
  {arXiv:1702.03163 [hep-th]} \BibitemShut {NoStop}%
\bibitem [{\citenamefont {Arkani-Hamed}\ and\ \citenamefont
  {Yuan}(2017)}]{Arkani-Hamed:2017ahv}%
  \BibitemOpen
  \bibfield  {author} {\bibinfo {author} {\bibfnamefont {N.}~\bibnamefont
  {Arkani-Hamed}}\ and\ \bibinfo {author} {\bibfnamefont {E.~Y.}\ \bibnamefont
  {Yuan}},\ }\href@noop {} {\  (\bibinfo {year} {2017})},\ \Eprint
  {http://arxiv.org/abs/1712.09991} {arXiv:1712.09991 [hep-th]} \BibitemShut
  {NoStop}%
\bibitem [{\citenamefont {Herrmann}\ and\ \citenamefont
  {Parra-Martinez}(2020)}]{Herrmann:2019upk}%
  \BibitemOpen
  \bibfield  {author} {\bibinfo {author} {\bibfnamefont {E.}~\bibnamefont
  {Herrmann}}\ and\ \bibinfo {author} {\bibfnamefont {J.}~\bibnamefont
  {Parra-Martinez}},\ }\href {\doibase 10.1007/JHEP02(2020)099} {\bibfield
  {journal} {\bibinfo  {journal} {JHEP}\ }\textbf {\bibinfo {volume} {02}},\
  \bibinfo {pages} {099} (\bibinfo {year} {2020})},\ \Eprint
  {http://arxiv.org/abs/1909.04777} {arXiv:1909.04777 [hep-th]} \BibitemShut
  {NoStop}%
\bibitem [{\citenamefont {Bourjaily}\ \emph
  {et~al.}(2020{\natexlab{b}})\citenamefont {Bourjaily}, \citenamefont {Gardi},
  \citenamefont {McLeod},\ and\ \citenamefont {Vergu}}]{Bourjaily:2019exo}%
  \BibitemOpen
  \bibfield  {author} {\bibinfo {author} {\bibfnamefont {J.~L.}\ \bibnamefont
  {Bourjaily}}, \bibinfo {author} {\bibfnamefont {E.}~\bibnamefont {Gardi}},
  \bibinfo {author} {\bibfnamefont {A.~J.}\ \bibnamefont {McLeod}}, \ and\
  \bibinfo {author} {\bibfnamefont {C.}~\bibnamefont {Vergu}},\ }\href
  {\doibase 10.1007/JHEP08(2020)029} {\bibfield  {journal} {\bibinfo  {journal}
  {JHEP}\ }\textbf {\bibinfo {volume} {08}},\ \bibinfo {pages} {029} (\bibinfo
  {year} {2020}{\natexlab{b}})},\ \Eprint {http://arxiv.org/abs/1912.11067}
  {arXiv:1912.11067 [hep-th]} \BibitemShut {NoStop}%
\bibitem [{Note2()}]{Note2}%
  \BibitemOpen
  \bibinfo {note} {From now on, we will omit the dependence on
  kinematics.}\BibitemShut {Stop}%
\bibitem [{\citenamefont {Li}\ and\ \citenamefont {Zhang}(2021)}]{Li:2021bwg}%
  \BibitemOpen
  \bibfield  {author} {\bibinfo {author} {\bibfnamefont {Z.}~\bibnamefont
  {Li}}\ and\ \bibinfo {author} {\bibfnamefont {C.}~\bibnamefont {Zhang}},\
  }\href {\doibase 10.1007/JHEP12(2021)113} {\bibfield  {journal} {\bibinfo
  {journal} {JHEP}\ }\textbf {\bibinfo {volume} {12}},\ \bibinfo {pages} {113}
  (\bibinfo {year} {2021})},\ \Eprint {http://arxiv.org/abs/2110.00350}
  {arXiv:2110.00350 [hep-th]} \BibitemShut {NoStop}%
\bibitem [{\citenamefont {Besier}\ \emph {et~al.}(2019)\citenamefont {Besier},
  \citenamefont {Van~Straten},\ and\ \citenamefont
  {Weinzierl}}]{Besier:2018jen}%
  \BibitemOpen
  \bibfield  {author} {\bibinfo {author} {\bibfnamefont {M.}~\bibnamefont
  {Besier}}, \bibinfo {author} {\bibfnamefont {D.}~\bibnamefont {Van~Straten}},
  \ and\ \bibinfo {author} {\bibfnamefont {S.}~\bibnamefont {Weinzierl}},\
  }\href {\doibase 10.4310/CNTP.2019.v13.n2.a1} {\bibfield  {journal} {\bibinfo
   {journal} {Commun. Num. Theor. Phys.}\ }\textbf {\bibinfo {volume} {13}},\
  \bibinfo {pages} {253} (\bibinfo {year} {2019})},\ \Eprint
  {http://arxiv.org/abs/1809.10983} {arXiv:1809.10983 [hep-th]} \BibitemShut
  {NoStop}%
\bibitem [{\citenamefont {Caron-Huot}\ and\ \citenamefont
  {Henn}(2014)}]{Caron-Huot:2014lda}%
  \BibitemOpen
  \bibfield  {author} {\bibinfo {author} {\bibfnamefont {S.}~\bibnamefont
  {Caron-Huot}}\ and\ \bibinfo {author} {\bibfnamefont {J.~M.}\ \bibnamefont
  {Henn}},\ }\href {\doibase 10.1007/JHEP06(2014)114} {\bibfield  {journal}
  {\bibinfo  {journal} {JHEP}\ }\textbf {\bibinfo {volume} {06}},\ \bibinfo
  {pages} {114} (\bibinfo {year} {2014})},\ \Eprint
  {http://arxiv.org/abs/1404.2922} {arXiv:1404.2922 [hep-th]} \BibitemShut
  {NoStop}%
\bibitem [{\citenamefont {Drummond}\ \emph {et~al.}(2011)\citenamefont
  {Drummond}, \citenamefont {Henn},\ and\ \citenamefont
  {Trnka}}]{Drummond:2010cz}%
  \BibitemOpen
  \bibfield  {author} {\bibinfo {author} {\bibfnamefont {J.~M.}\ \bibnamefont
  {Drummond}}, \bibinfo {author} {\bibfnamefont {J.~M.}\ \bibnamefont {Henn}},
  \ and\ \bibinfo {author} {\bibfnamefont {J.}~\bibnamefont {Trnka}},\ }\href
  {\doibase 10.1007/JHEP04(2011)083} {\bibfield  {journal} {\bibinfo  {journal}
  {JHEP}\ }\textbf {\bibinfo {volume} {04}},\ \bibinfo {pages} {083} (\bibinfo
  {year} {2011})},\ \Eprint {http://arxiv.org/abs/1010.3679} {arXiv:1010.3679
  [hep-th]} \BibitemShut {NoStop}%
\bibitem [{Note3()}]{Note3}%
  \BibitemOpen
  \bibinfo {note} {The bi-rational transformation $(s,y)\DOTSB \mapstochar
  \rightarrow (S,Y)$ such that $Y^2=4S^3-g_2S-g_3$ maps $(0,\protect \sqrt
  {-\protect \mathcal Q(0)})$ to $Y=0$ and $(\infty ,\infty )$ to $Y=\infty $.
  Since $(S,Y)=(\wp ,\wp ')$ is an isomorphism between the elliptic curve and
  the torus $\protect \mathbb {C}/\langle 1,\tau \rangle $, we see that
  $w(0)=\wp '^{-1}(0)=0$ mod $\langle 1,\tau \rangle /2$ and $w(\infty )=\wp
  '^{-1}(\infty )=0$ mod $\langle 1,\tau \rangle $.}\BibitemShut {Stop}%
\bibitem [{\citenamefont {Caron-Huot}\ and\ \citenamefont
  {Huang}(2013)}]{Caron-Huot:2012sos}%
  \BibitemOpen
  \bibfield  {author} {\bibinfo {author} {\bibfnamefont {S.}~\bibnamefont
  {Caron-Huot}}\ and\ \bibinfo {author} {\bibfnamefont {Y.-t.}\ \bibnamefont
  {Huang}},\ }\href {\doibase 10.1007/JHEP03(2013)075} {\bibfield  {journal}
  {\bibinfo  {journal} {JHEP}\ }\textbf {\bibinfo {volume} {03}},\ \bibinfo
  {pages} {075} (\bibinfo {year} {2013})},\ \Eprint
  {http://arxiv.org/abs/1210.4226} {arXiv:1210.4226 [hep-th]} \BibitemShut
  {NoStop}%
\bibitem [{\citenamefont {He}\ \emph {et~al.}(2023{\natexlab{a}})\citenamefont
  {He}, \citenamefont {Huang}, \citenamefont {Kuo},\ and\ \citenamefont
  {Li}}]{He:2022lfz}%
  \BibitemOpen
  \bibfield  {author} {\bibinfo {author} {\bibfnamefont {S.}~\bibnamefont
  {He}}, \bibinfo {author} {\bibfnamefont {Y.-t.}\ \bibnamefont {Huang}},
  \bibinfo {author} {\bibfnamefont {C.-K.}\ \bibnamefont {Kuo}}, \ and\
  \bibinfo {author} {\bibfnamefont {Z.}~\bibnamefont {Li}},\ }\href {\doibase
  10.1007/JHEP02(2023)065} {\bibfield  {journal} {\bibinfo  {journal} {JHEP}\
  }\textbf {\bibinfo {volume} {02}},\ \bibinfo {pages} {065} (\bibinfo {year}
  {2023}{\natexlab{a}})},\ \Eprint {http://arxiv.org/abs/2211.01792}
  {arXiv:2211.01792 [hep-th]} \BibitemShut {NoStop}%
\bibitem [{\citenamefont {He}\ \emph {et~al.}(2023{\natexlab{b}})\citenamefont
  {He}, \citenamefont {Kuo}, \citenamefont {Li},\ and\ \citenamefont
  {Zhang}}]{He:2023exb}%
  \BibitemOpen
  \bibfield  {author} {\bibinfo {author} {\bibfnamefont {S.}~\bibnamefont
  {He}}, \bibinfo {author} {\bibfnamefont {C.-K.}\ \bibnamefont {Kuo}},
  \bibinfo {author} {\bibfnamefont {Z.}~\bibnamefont {Li}}, \ and\ \bibinfo
  {author} {\bibfnamefont {Y.-Q.}\ \bibnamefont {Zhang}},\ }\href@noop {} {\
  (\bibinfo {year} {2023}{\natexlab{b}})},\ \Eprint
  {http://arxiv.org/abs/2303.03035} {arXiv:2303.03035 [hep-th]} \BibitemShut
  {NoStop}%
\bibitem [{\citenamefont {Henn}\ \emph {et~al.}(2023)\citenamefont {Henn},
  \citenamefont {Lagares},\ and\ \citenamefont {Zhang}}]{Henn:2023pkc}%
  \BibitemOpen
  \bibfield  {author} {\bibinfo {author} {\bibfnamefont {J.~M.}\ \bibnamefont
  {Henn}}, \bibinfo {author} {\bibfnamefont {M.}~\bibnamefont {Lagares}}, \
  and\ \bibinfo {author} {\bibfnamefont {S.-Q.}\ \bibnamefont {Zhang}},\
  }\href@noop {} {\  (\bibinfo {year} {2023})},\ \Eprint
  {http://arxiv.org/abs/2303.02996} {arXiv:2303.02996 [hep-th]} \BibitemShut
  {NoStop}%
\bibitem [{Note4()}]{Note4}%
  \BibitemOpen
  \bibinfo {note} {To show that $t=-1$ is a root, simply notice that when
  $t=-1$, the first three rows/columns are linearly dependent.}\BibitemShut
  {Stop}%
\bibitem [{\citenamefont {Cao}\ \emph {et~al.}(2023)\citenamefont {Cao},
  \citenamefont {He},\ and\ \citenamefont {Tang}}]{Cao:2023tpx}%
  \BibitemOpen
  \bibfield  {author} {\bibinfo {author} {\bibfnamefont {Q.}~\bibnamefont
  {Cao}}, \bibinfo {author} {\bibfnamefont {S.}~\bibnamefont {He}}, \ and\
  \bibinfo {author} {\bibfnamefont {Y.}~\bibnamefont {Tang}},\ }\href@noop {}
  {\  (\bibinfo {year} {2023})},\ \Eprint {http://arxiv.org/abs/2301.07834}
  {arXiv:2301.07834 [hep-th]} \BibitemShut {NoStop}%
\bibitem [{\citenamefont {McLeod}\ \emph {et~al.}(2023)\citenamefont {McLeod},
  \citenamefont {Morales}, \citenamefont {von Hippel}, \citenamefont
  {Wilhelm},\ and\ \citenamefont {Zhang}}]{McLeod:2023qdf}%
  \BibitemOpen
  \bibfield  {author} {\bibinfo {author} {\bibfnamefont {A.}~\bibnamefont
  {McLeod}}, \bibinfo {author} {\bibfnamefont {R.}~\bibnamefont {Morales}},
  \bibinfo {author} {\bibfnamefont {M.}~\bibnamefont {von Hippel}}, \bibinfo
  {author} {\bibfnamefont {M.}~\bibnamefont {Wilhelm}}, \ and\ \bibinfo
  {author} {\bibfnamefont {C.}~\bibnamefont {Zhang}},\ }\href@noop {} {\
  (\bibinfo {year} {2023})},\ \Eprint {http://arxiv.org/abs/2301.07965}
  {arXiv:2301.07965 [hep-th]} \BibitemShut {NoStop}%
\bibitem [{\citenamefont {Yang}(2022)}]{Yang:2022gko}%
  \BibitemOpen
  \bibfield  {author} {\bibinfo {author} {\bibfnamefont {Q.}~\bibnamefont
  {Yang}},\ }\href {\doibase 10.1007/JHEP08(2022)168} {\bibfield  {journal}
  {\bibinfo  {journal} {JHEP}\ }\textbf {\bibinfo {volume} {08}},\ \bibinfo
  {pages} {168} (\bibinfo {year} {2022})},\ \Eprint
  {http://arxiv.org/abs/2203.16112} {arXiv:2203.16112 [hep-th]} \BibitemShut
  {NoStop}%
\bibitem [{\citenamefont {He}\ \emph {et~al.}(2022{\natexlab{b}})\citenamefont
  {He}, \citenamefont {Liu}, \citenamefont {Tang},\ and\ \citenamefont
  {Yang}}]{He:2022tph}%
  \BibitemOpen
  \bibfield  {author} {\bibinfo {author} {\bibfnamefont {S.}~\bibnamefont
  {He}}, \bibinfo {author} {\bibfnamefont {J.}~\bibnamefont {Liu}}, \bibinfo
  {author} {\bibfnamefont {Y.}~\bibnamefont {Tang}}, \ and\ \bibinfo {author}
  {\bibfnamefont {Q.}~\bibnamefont {Yang}},\ }\href@noop {} {\  (\bibinfo
  {year} {2022}{\natexlab{b}})},\ \Eprint {http://arxiv.org/abs/2207.13482}
  {arXiv:2207.13482 [hep-th]} \BibitemShut {NoStop}%
\bibitem [{\citenamefont {Abreu}\ \emph
  {et~al.}(2017{\natexlab{b}})\citenamefont {Abreu}, \citenamefont {Britto},
  \citenamefont {Duhr},\ and\ \citenamefont {Gardi}}]{Abreu:2017enx}%
  \BibitemOpen
  \bibfield  {author} {\bibinfo {author} {\bibfnamefont {S.}~\bibnamefont
  {Abreu}}, \bibinfo {author} {\bibfnamefont {R.}~\bibnamefont {Britto}},
  \bibinfo {author} {\bibfnamefont {C.}~\bibnamefont {Duhr}}, \ and\ \bibinfo
  {author} {\bibfnamefont {E.}~\bibnamefont {Gardi}},\ }\href {\doibase
  10.1103/PhysRevLett.119.051601} {\bibfield  {journal} {\bibinfo  {journal}
  {Phys. Rev. Lett.}\ }\textbf {\bibinfo {volume} {119}},\ \bibinfo {pages}
  {051601} (\bibinfo {year} {2017}{\natexlab{b}})},\ \Eprint
  {http://arxiv.org/abs/1703.05064} {arXiv:1703.05064 [hep-th]} \BibitemShut
  {NoStop}%
\bibitem [{\citenamefont {Abreu}\ \emph
  {et~al.}(2017{\natexlab{c}})\citenamefont {Abreu}, \citenamefont {Britto},
  \citenamefont {Duhr},\ and\ \citenamefont {Gardi}}]{Abreu:2017mtm}%
  \BibitemOpen
  \bibfield  {author} {\bibinfo {author} {\bibfnamefont {S.}~\bibnamefont
  {Abreu}}, \bibinfo {author} {\bibfnamefont {R.}~\bibnamefont {Britto}},
  \bibinfo {author} {\bibfnamefont {C.}~\bibnamefont {Duhr}}, \ and\ \bibinfo
  {author} {\bibfnamefont {E.}~\bibnamefont {Gardi}},\ }\href {\doibase
  10.1007/JHEP12(2017)090} {\bibfield  {journal} {\bibinfo  {journal} {JHEP}\
  }\textbf {\bibinfo {volume} {12}},\ \bibinfo {pages} {090} (\bibinfo {year}
  {2017}{\natexlab{c}})},\ \Eprint {http://arxiv.org/abs/1704.07931}
  {arXiv:1704.07931 [hep-th]} \BibitemShut {NoStop}%
\bibitem [{\citenamefont {Abreu}\ \emph {et~al.}(2020)\citenamefont {Abreu},
  \citenamefont {Britto}, \citenamefont {Duhr}, \citenamefont {Gardi},\ and\
  \citenamefont {Matthew}}]{Abreu:2019eyg}%
  \BibitemOpen
  \bibfield  {author} {\bibinfo {author} {\bibfnamefont {S.}~\bibnamefont
  {Abreu}}, \bibinfo {author} {\bibfnamefont {R.}~\bibnamefont {Britto}},
  \bibinfo {author} {\bibfnamefont {C.}~\bibnamefont {Duhr}}, \bibinfo {author}
  {\bibfnamefont {E.}~\bibnamefont {Gardi}}, \ and\ \bibinfo {author}
  {\bibfnamefont {J.}~\bibnamefont {Matthew}},\ }\href {\doibase
  10.22323/1.375.0065} {\bibfield  {journal} {\bibinfo  {journal} {PoS}\
  }\textbf {\bibinfo {volume} {RADCOR2019}},\ \bibinfo {pages} {065} (\bibinfo
  {year} {2020})},\ \Eprint {http://arxiv.org/abs/1912.06561} {arXiv:1912.06561
  [hep-th]} \BibitemShut {NoStop}%
\bibitem [{\citenamefont {Abreu}\ \emph {et~al.}(2021)\citenamefont {Abreu},
  \citenamefont {Britto}, \citenamefont {Duhr}, \citenamefont {Gardi},\ and\
  \citenamefont {Matthew}}]{Abreu:2021vhb}%
  \BibitemOpen
  \bibfield  {author} {\bibinfo {author} {\bibfnamefont {S.}~\bibnamefont
  {Abreu}}, \bibinfo {author} {\bibfnamefont {R.}~\bibnamefont {Britto}},
  \bibinfo {author} {\bibfnamefont {C.}~\bibnamefont {Duhr}}, \bibinfo {author}
  {\bibfnamefont {E.}~\bibnamefont {Gardi}}, \ and\ \bibinfo {author}
  {\bibfnamefont {J.}~\bibnamefont {Matthew}},\ }\href {\doibase
  10.1007/JHEP10(2021)131} {\bibfield  {journal} {\bibinfo  {journal} {JHEP}\
  }\textbf {\bibinfo {volume} {10}},\ \bibinfo {pages} {131} (\bibinfo {year}
  {2021})},\ \Eprint {http://arxiv.org/abs/2106.01280} {arXiv:2106.01280
  [hep-th]} \BibitemShut {NoStop}%
\end{thebibliography}%

\end{document}